\documentclass[11pt]{article}
\usepackage{graphicx}
\usepackage{amsmath,amssymb,amsthm,amsfonts}
\usepackage{mathrsfs}
\usepackage{color}
\usepackage{amssymb}
\usepackage{cite}
\usepackage{soul}
\newtheorem{thm}{Theorem}[section]

\newtheorem{lem}[thm]{Lemma}

\theoremstyle{definition}
\newtheorem{defn}{Definition}[section]
\theoremstyle{remark}
\usepackage{appendix}
\newtheorem{rem}{Remark}[section]

\numberwithin{equation}{section}

\DeclareMathSymbol{\C}{\mathalpha}{AMSb}{"43}

\newcommand{\lam}{\lambda}

\newcommand{\R}{{\mathbb{R}}}

\newcommand{\inte}{\int_{\mathbb{R}^2}}

\def\R{{\mathbb R}}
\def\C{{\mathbb C}}
\def\inte{\int_{\mathbb{R}^3}}
\def\f{\frac}
\def\al{\alpha}
\def\ga{\gamma}
\def\va{\varepsilon}
\def\vaa{\varepsilon_a}
\def\E{\mathcal{E}}

\def\l{\left}
\def\r{\right}
\def\la{\langle}
\def\ra{\rangle}
\def\ni{\noindent}

\newcommand{\bsub}{\begin{subequations}}
\newcommand{\esub}{\end{subequations}$\!$}

\begin{document}
\title{ Mass Concentration of Two-Spinless Fermi  Systems with Attractive Interactions}
\author{Yujin Guo\thanks{Email: yguo@ccnu.edu.cn. Y. J. Guo is partially supported by National Key R $\&$ D Program of China (Grant 2023YFA1010001), and NSF of China (Grants 12225106 and 11931012).}\ \ \, and\ \ Yan Li\thanks{Email: yanlimath@mails.ccnu.edu.cn.}\\
\small \it	School of Mathematics and Statistics,\\
\small \it  Key Laboratory of Nonlinear Analysis {\em \&} Applications (Ministry of Education),\\
\small \it Central China Normal University, Wuhan 430079, P. R. China\\
}

\date{\today}

\smallbreak \maketitle

\begin{abstract}
We study the  two-spinless mass-critical Fermi systems with attractive interactions and trapping potentials. We prove that ground states of the system exist, if and only if the  strength $a$ of attractive interactions satisfies $0<a<a_2^*$, where $0<a_2^*<+\infty$ is the best constant of a dual finite-rank Lieb-Thirring inequality. By the blow-up analysis of many-fermion systems, we show that ground states of the system  concentrate at the flattest minimum points of the trapping potential $V(x)$ as $a\nearrow a_2^*$.
\end{abstract}

\vskip 0.05truein

\noindent {\it Keywords:} Fermi systems; ground states; mass concentration

\vskip 0.2truein

\section{Introduction}
Over the past few decades, experimental achievements of trapped atomic gases have revealed (cf.  \cite{G-S,Blo,Chin,Blo1}) the beautiful and subtle physics of the quantum world for ultracold atoms.
These experiments were usually carried out in the presence of optical laser traps that confine the particles in a limited region of the space, see  \cite{Blo1,Met}.  In particular, spinless fermions in  harmonic traps have played a crucial role of  recent developments  (cf. \cite{Bra,G-S,Vicari}),
given that trapping potentials in many experiments can be safely approximated with the harmonic form.  Moreover,
when spinless Fermi gases are confined in inhomogeneous traps \cite{Rad}, the nonuniform density leads to the spatially varying energy and length scales. We also refer the reader to
\cite{Muk} for creating homogeneous Fermi gases of ultracold atoms in a uniform potential. These experiments have   generated some interesting theoretical questions. Numerical simulations and mathematical theories of trapped fermions  have therefore been a focus of research interests in physics and mathematics since the last decades (cf. \cite{G-S,Naf,Dav,Dou,Bra,Lieb1}).


 Following the arguments of  \cite{Frank2,Lieb1,Lewin},  ground states of two-spinless mass-critical  Fermi  systems with attractive interactions and trapping potentials can be described by the minimizers of  the following constraint variational problem:
\begin{equation}\label{0.1}
\begin{aligned}
  E_a(2):&=\inf
  \Big\{\E_a(\Psi):\ \|\Psi\|_2^2=1,\ \Psi\in\wedge^2 L^2(\R^{3},\C)\cap H^1(\R^{6},\C),\\
  &\qquad\qquad \qquad\sum_{i=1}^2\int_{\R^{6}}V(x_i)|\Psi|^2dx_1dx_2<\infty\Big\},\ \ a>0,
\end{aligned}
\end{equation}
where the energy functional $\E_a(\Psi)$  satisfies
\begin{equation*}
  \E_a(\Psi):=\sum\limits_{i=1}^2\int_{\R^{6}}\Big(|\nabla_{x_i}\Psi|^2
  +V(x_i)|\Psi|^2\Big)dx_1dx_2-a\inte\rho_{\Psi}^{\f53}(x)dx.
\end{equation*}
Here $\wedge^2 L^2(\R^{3},\C)$ is the subspace of $L^2(\R^{6},\C)$ consisting of  all antisymmetric wave functions, $V(x)\geq 0$ denotes the trapping  potential, $a>0$ represents the attractive strength of the quantum particles, and the one-particle density $\rho_\Psi$ associated with $\Psi$ is defined by
\begin{equation*}
  \rho_\Psi(x):=2\int_{\R^{3}}|\Psi(x,x_2)|^2dx_2.
\end{equation*}

Applying the approach of  \cite[Appendix A
and Lemma 2.3]{Chen}, the problem \eqref{0.1} can be reduced equivalently to the following form
\begin{equation}\label{0.5}
\begin{aligned}
  E_a(2)&=\inf\Big\{\E_a(\ga):\ \ga=\sum\limits_{i=1}^2|u_i\ra\la u_i|,\ u_i\in \mathcal{H},\\
  &\qquad\qquad\qquad(u_i,u_j)=\delta_{ij},\ i,j=1,2\Big\},\ \ a>0,
\end{aligned}
\end{equation}
where the energy functional $\E_a(\ga)$ satisfies
\begin{equation}\label{0.6}
  \E_a(\ga):=\mathrm{Tr}\big(-\Delta+V(x)\big)\ga-a\inte\rho_\ga^{\f53}(x)dx,
\end{equation}
and the Hilbert space $\mathcal{H}$ is defined by
\begin{equation*}
  \mathcal{H}:=\Big\{u\in H^1(\R^3,\R):\ \inte V(x)|u(x)|^2dx<\infty\Big\}.
\end{equation*}
Here  the non-negative self-adjoint operator
$
  \ga=\sum_{i=1}^2|u_i\ra\la u_i|
$
on $L^2(\R^3,\R)$ satisfies
$$\ga\varphi(x)=\sum_{i=1}^2u_i(x)(\varphi,u_i)_{L^2(\R^3,\R)},\ \ \forall\varphi\in L^2(\R^3,\R),$$
the kinetic energy of $\ga$ is denoted by
\begin{equation}\label{1-3}
  \mathrm{Tr}(-\Delta\ga):=\sum_{j=1}^3 \mathrm{Tr}(P_j\ga P_j)=\sum_{j=1}^3\sum_{i=1}^2\|P_j u_i\|_{L^2}^2=\sum_{i=1}^2\inte|\nabla u_i(x)|^2dx,
\end{equation}
where $P_j:=-i\partial_{x_j}$, and  the corresponding density of $\ga$ is defined as
\begin{equation}\label{0.8}
  \rho_{\ga}(x):=\sum_{i=1}^2|u_i(x)|^2.
\end{equation}
%

If the trapping  potential $V(x)$ in \eqref{0.6} is ignored, the existence of minimizers for $E_a(2)$ in the $L^2$-subcritical case was analyzed in  \cite{Lewin}. Motivated by \cite{Lewin}, the authors in \cite{Chen} studied the existence and concentration behavior of minimizers for $E_a(2)$ in the $L^2$-subcritical case, where $V(x)<0$ is the Coulomb potential. Further, the  $L^2$-critical case of $E_a(2)$ with the Coulomb potential was recently  considered in \cite{Chen1}.  On the other hand, the physical experiments of Fermi gases were also performed in other types of trapping potentials over the past few years, such as harmonic potentials, double-well potentials, and so on (cf. \cite{Bra,G-S,Vicari,Naf}).
Moreover, once the problem $E_a(2)$ is analyzed with other types of traps, instead of the Coulomb form, some extra difficulties appear especially in the analysis of the Lagrange multipliers for $E_a(2)$.
Inspired by above facts, the purpose of the present paper is to study the problem $E_a(2)$ with the trap $0\leq V(x)\in L_{loc}^\infty(\R^3)$ satisfying $\lim_{|x|\to\infty}V(x)=\infty$.

The existing investigations (cf. \cite{Chen1}) show that the problem $E_a(2)$ is related to the
following minimization  problem
\begin{equation}\label{0.7}
\begin{aligned}
    0<a_2^*:=&\inf\Big\{\f{\|\ga\|^{\f23}\mathrm{Tr}(-\Delta\ga)}{\inte\rho_\ga^{\f53}(x)dx}: \ 0\leq \ga=\ga^*,\  \hbox{Rank}(\ga)\leq2
    \Big\}.
\end{aligned}
\end{equation}
Here $\ga$ is of the form $\ga=\sum_{i=1}^2n_i|u_i\ra\la u_i|$, where $n_i\geq 0$ and $u_i\in H^1(\R^3)$ satisfies $(u_i,u_j)=\delta_{ij}$ for $i,j=1,2$, $\rho_\ga(x)$ is defined as  $\rho_\ga(x)=\sum_{i=1}^2n_i u_i^2(x)$, and $\|\ga\|$ is the operator norm. Note from \cite[Theorem 6]{Frank2} that  the problem $a_2^*$ defined in \eqref{0.7} admits at least one  minimizer. Moreover, any minimizer $\ga^{(2)}$ of the problem $a_2^*$ has rank $2$, and  can be written in the form
\begin{equation*}
  \ga^{(2)}=\|\ga^{(2)}\|\sum\limits_{i=1}^{2}|Q_i\ra\la Q_i|,\ \ Q_i\in H^1(\R^3),\ \  (Q_i,Q_j)=\delta_{ij},\ \ i,j=1,2,
\end{equation*}
where the orthonormal system $(Q_1,Q_2)$ satisfies the following nonlinear Schr\"odinger system
\begin{equation}\label{0.10}
  \Big[-\Delta-\f{5a_2^*}3 \Big(\sum\limits_{j=1}^{2}Q_j^2\Big)^{\f23}\Big]Q_i=\hat\mu_iQ_i\ \ \hbox{in}\ \ \R^3,\ \ i=1,2,
\end{equation}
 and $\hat\mu_1<\hat\mu_2<0$ are the 2-first negative eigenvalues of the operator
$-\Delta-\f{5a_2^*}3 \Big(\sum\limits_{j=1}^{2}Q_j^2\Big)^{\f23}$ in $\R^3$.


Associated to the problem $E_a(2)$, we now define ground states of a fermionic nonlinear Schr\"odinger system, in the following sense that

\begin{defn}
({\em Ground states}). Suppose $0\leq V(x)\in L_{loc}^\infty(\R^3)$ satisfies $\lim_{|x|\to\infty}V(x)=\infty$.  A system $(u_1,u_2)\in \l(H^1(\R^3)\r)^2$, where $(u_i,u_j)=\delta_{ij}$ holds for $i,j=1,2$, is called a ground state of
\begin{equation}\label{0.11}
  H_Vu_i:=\Big[-\Delta+V(x)-\f{5a}{3}\Big(\sum\limits_{j=1}^2 u_j^2\Big)^{\f23}\Big]u_i=\mu_i u_i\ \ \hbox{in}\ \ \R^3,\ \ i=1,2,\ \ a>0,
\end{equation}
if it satisfies the system \eqref{0.11}, where $\mu_1<\mu_2$ are the 2-first eigenvalues  of the operator $H_V$ in $\R^3$.
\end{defn}

The first result of the present paper is concerned with the following existence of minimizers for $E_a(2)$ defined in \eqref{0.5}.

\begin{thm}\label{thm1}
Let $a_2^*>0$ be defined by \eqref{0.7}, and assume the potential $0\le V(x)\in L_{loc}^\infty(\R^3)$ satisfies $\lim_{|x|\to\infty}V(x)=\infty$. Then we have
\begin{enumerate}
\item If $0<a<a_2^*$, then there exists at least one minimizer $\ga=\sum_{i=1}^2|u_i\ra\la u_i|$ of $E_a(2)$, where $(u_1,u_2)$  is a ground state of  \eqref{0.11}.
\item If $a\geq a_2^*$, then there is no minimizer of $E_a(2)$.
\end{enumerate}
\end{thm}

When the Coulomb potential $V(x)<0$ is considered, the  non-existence of minimizers for $E_{a}(2)$ is proved in \cite{Chen1}, which gives that $E_{a}(2)=-\infty$ for $a\geq a_2^*$, by applying the properties of the Coulomb potential and
the monotonicity of the energy $E_a(2)$ with respect to the parameter $a>0$.
Different from \cite{Chen1}, we shall however derive that $E_{a_2^*}(2)=0$ and $E_a(2)=-\infty$ for $a>a_2^*$ by constructing suitable orthogonal test functions. The  non-existence result of $E_{a_2^*}(2)$ is further proved by applying the properties of  minimizers for the problem $a_2^*$ defined in \eqref{0.7}.

On the other hand, the existence of Theorem \ref{thm1} is derived by analyzing the compactness of the minimizing sequences, which can actually be extended to the problem $E_a(N)$ with any $N\in \mathbb{N}^+$. Moreover, we shall prove, in a simplifier way than those of \cite{Lewin,Chen,Chen1}, that the minimizers of $E_a(2)$ are essentially ground states of  \eqref{0.11}.
Further, assume $\ga_a=\sum_{i=1}^2|u_i^a\ra\la u_i^a|$ is a minimizer of  $E_a(2)$ for $0<a<a_2^*$, then the proof of Theorem \ref{thm1} yields that $\inte V(x)\rho_{\ga_a}(x)dx\to \inf_{x\in\R^3}V(x)$ as $a\nearrow a_2^*$, which implies roughly that the mass of the minimizers $\ga_a$ concentrates at the global minimum points of $V(x)$  as $a\nearrow a_2^*$.  The main purpose of the present paper is to further analyze the mass concentration behavior of the minimizers $\ga_a$ as $a\nearrow a_2^*$.

Towards the above main purpose, we now assume that there exist positive constants $p_1,\cdots, p_l$ and $C$ such that
\begin{equation}\label{0-2}
  V(x)=g(x)\prod_{m=1}^l|x-x_m|^{p_m}\ \ \hbox{and}\ \ C< g(x)<\f1C\ \ \hbox{in}\ \ \R^3,
\end{equation}
where $x_m\neq x_n$ for $m\neq n$,  $g(x)\in C_{loc}^\kappa(\R^3)$ for some $\kappa\in(0,1)$, and the limits $\lim\limits_{x\to x_m}g(x)$ exist for all $1\leq m\leq l$.
Denote
\begin{equation}\label{0-4}
  p=\max\{p_1,\cdots,p_l\}>0,\ \ \Lambda:=\{x\in\R^3:\, V(x)=0\}=\{x_1,\cdots,x_l\},
\end{equation}
and
\begin{equation}\label{0-3}
  \mathcal{Z}:=\{x_m\in\Lambda:\ \al_m=\al\},
\end{equation}
where
\begin{equation}\label{0-5}
  \al:=\min_{1\leq m\leq l}\big\{\al_m\big\}>0,\ \ \hbox{and}\ \  \al_m=\lim\limits_{x\to x_m}\f{V(x)}{|x-x_m|^p}\in(0,+\infty].
\end{equation}
Note from (\ref{0-3}) that the set $\mathcal{Z}$ denotes the locations of the flattest global minimum points for $V(x)$. We remark that \eqref{0-2} covers both the harmonic trap and double-well trap, which were achieved experimentally in \cite{Bra,G-S,Vicari,Naf}.

Using above notations, the main result of the present paper can be stated as the following theorem:

\begin{thm}\label{thm3}
Suppose $V(x)$ satisfies \eqref{0-2}, and let $\ga_a=\sum_{i=1}^2|u_i^a\ra\la u_i^a|$ be a minimizer of $E_a(2)$ for $0<a<a_2^*$, where $u_i^a$ satisfies \eqref{0.11} for $i=1,2$. Then for any given sequence $\{a_n\}$ with $a_n\nearrow a_2^*$ as $n\to\infty$, there exists a subsequence, still denoted by $\{a_n\}$, of $\{a_n\}$  such that for $i=1,2$,
\begin{equation}\label{1-2}
\begin{aligned}
  w_{i}^{a_n}(x):&=(a_2^*-a_n)^{\f{3}{2(p+2)}}u_i^{a_n}\Big((a_2^*-a_n)^{\f{1}{p+2}}x+x_{a_n}\Big)\\
  &\qquad\to w_{i}(x)\ \ \hbox{strongly in}\ \ H^1(\R^3)\cap L^\infty(\R^3)\ \ \hbox{as}\ \ n\to\infty,
\end{aligned}
\end{equation}
where $p>0$ is as in \eqref{0-4}, $\ga:=\sum_{i=1}^2|w_{i}\ra\la w_{i}|$ is a minimizer of $a_2^*$, and the global maximum point  $x_{a_n}$ of the density $\rho_{\ga_{a_n}}(x)=\sum_{i=1}^2|u_i^{a_n}|^2$ satisfies
\begin{equation}\label{1-2M}
  \lim\limits_{n\to\infty}\f{x_{a_n}-x_k}{(a_2^*-a_n)^{\f{1}{p+2}}}=\bar x
\end{equation}
for some points $x_k\in\mathcal{Z}$ and $\bar x\in\R^3$.

\end{thm}
\begin{rem}
(1). It follows from Theorem \ref{thm3} that the minimizers of $E_{a_n}(2)$ concentrate at the flattest minimum points of $V(x)$ as $a_n\nearrow a_2^*$.

(2). Under the assumption \eqref{0-2}, Theorem \ref{thm3} yields that the minimizer $\ga_{a_n}=\sum_{i=1}^2|u_i^{a_n}\ra\la u_i^{a_n}|$, where $u_i^{a_n}$ satisfies \eqref{0.11} for $i=1,2$, of $E_{a_n}(2)$ behaves like
\begin{equation*}
  \ga_{a_n}(x,y)\approx (a_2^*-a_n)^{-\f{3}{p+2}}\ga\Big(\f{x-x_{a_n}}{(a_2^*-a_n)^{\f{1}{p+2}}},\f{y-x_{a_n}}{(a_2^*-a_n)^{\f{1}{p+2}}}\Big)\ \ \hbox{as}\ \ a_n\nearrow a_2^*,
\end{equation*}
where $\ga(x,y)=\sum_{i=1}^2w_{i}(x)w_{i}(y)$ is the integral kernel of $\ga$, and  the energy $E_{a_n}(2)$ satisfies
\begin{equation*}
  \lim\limits_{a_n\nearrow a_2^*}\frac{E_{a_n}(2)}{(a_2^*-a_n)^{\f{p}{p+2}}}
  =\inte\rho_{\ga}^{\f53}(x)dx+\al\inte|x+\bar x|^p\rho_{\ga}(x)dx,
\end{equation*}
where $\al>0$ is defined by \eqref{0-5}.
\end{rem}

There are several further comments on Theorem \ref{thm3} which is proved by the  blow-up analysis of many-body fermions. Firstly, comparing with the existing results of \cite{Chen1}, Theorem \ref{thm3} can provide additionally the refined information on the maximum point of the density $\rho_{\ga_{a_n}}(x)$ as $a_n\nearrow a_2^*$. Secondly, the argument of \cite{Chen,Chen1} is improved  to obtain the $H^1$-convergence (\ref{1-2}) of Theorem \ref{thm3}. Thirdly, the proof of Theorem \ref{thm3}  needs the following estimates:
\begin{equation}\label{0-1}
  \mu_1^{a_n}<\mu_2^{a_n}<0\ \ \hbox{as}\ \ a_n\nearrow a_2^*,
\end{equation}
where $\mu_1^{a_n}$ and $\mu_2^{a_n}$ are the 2-first eigenvalues of the operator $-\Delta+V(x)-\f{5a_n}{3}\rho_{\ga_{a_n}}^{2/3}$ in $\R^3$. We shall derive \eqref{0-1} in Section 4 by the refined analysis of the energy $E_{a_n}(2)$.

This paper is organized as follows. Section 2 is devoted to the proof of Theorem \ref{thm1} on the existence and non-existence of minimizers for $E_a(2)$. We analyze in Section 3 some refined estimates of minimizers for $E_a(2)$, based on which  the proof of Theorem \ref{thm3} is given in Section 4. The  exponential decay of minimizers for $a_2^*$ is finally addressed in Appendix A.

\section{Existence and Non-existence of Minimizers}
In this section, we shall establish Theorem \ref{thm1} on the existence and non-existence of minimizers for $E_a(2)$ defined by \eqref{0.5}.
Towards this purpose, we first recall  the following compactness result (see e.g. \cite[Theorem XIII.67]{Reed1} or \cite{Adams}):

\begin{lem}\label{lem2.1}
Suppose $0\le V(x)\in L^\infty_{loc}(\R^3)$ satisfies $\lim_{|x|\to\infty}V(x)=\infty$.
Then for any $2\leq q< 6$, the embedding  $\mathcal{H}\hookrightarrow L^q(\R^3)$  is compact.
\end{lem}

Employing Lemma \ref{lem2.1}, we next  complete the proof of Theorem \ref{thm1}.
\vskip 0.05truein

\ni{\bf Proof of Theorem \ref{thm1}.} Without loss of generality, we assume additionally that the potential $V(x)\ge 0$ satisfies $\inf_{x\in\R^3}V(x)=0$.

1. We first prove the existence of minimizers for  $E_a(2)$, where $0<a<a_2^*$.
Let $\ga=\sum_{i=1}^2|u_i\ra\la u_i|$ be an operator satisfying $u_i\in\mathcal{H}$ and $(u_i,u_j)=\delta_{ij}$ for $i,j=1,2$. Since $V(x)\geq 0$, we obtain from \eqref{0.7} that for any $0< a<a_2^*$,
\begin{equation}\label{2.1}
\begin{aligned}
  \E_a(\ga)&=\mathrm{Tr}\big(-\Delta+V(x)\big)\ga-a\inte\rho_\ga^{\f53}(x)dx\\
  &\geq\l(1-\f{a}{a_2^*}\r)\mathrm{Tr}(-\Delta\ga)+\inte V(x)\rho_\ga(x)dx\\
  &\geq\l(1-\f{a}{a_2^*}\r)\mathrm{Tr}(-\Delta\ga)\geq 0,
\end{aligned}
\end{equation}
due to the fact that $\|\ga\|=1$.
This gives that $E_a(2)$ is bounded from below for $0< a<a_2^*$.

Let $\{\ga_n\}$ be a minimizing sequence of $E_a(2)$ satisfying  $\ga_n=\sum_{i=1}^2|u_i^n\ra\la u_i^n|$, $u_i^n\in\mathcal{H}$, $(u_i^n,u_j^n)=\delta_{ij}$ for $i,j=1,2$, and $\lim_{n\to\infty}\E_a(\ga_n)=E_a(2)$. We derive from  \eqref{2.1} that $\{u_i^n\}$ is bounded uniformly in $\mathcal{H}$ for $i=1,2$. Following Lemma \ref{2.1}, we obtain that there exists a function $u_i(x)\in\mathcal{H}$ such that for $i=1,2$,
\begin{equation*}
  u_i^n\rightharpoonup u_i\ \ \hbox{weakly in}\ \ \mathcal{H}\ \ \hbox{and}\ \ u_i^n\to u_i\ \ \hbox{strongly in}\ \ L^q(\R^3)\ \ \hbox{as}\ \ n\to\infty,\ \ 2\leq q<6.
\end{equation*}
Therefore, we have
\begin{equation*}
  (u_i,u_j)=\delta_{ij},\ \ i,j=1,2,
\end{equation*}
and
\begin{equation}\label{2.3}
  \rho_{\ga_n}
  =\sum\limits_{i=1}^2|u_i^n|^2\to\rho_\ga=\sum\limits_{i=1}^2|u_i|^2\ \ \hbox{strongly in}\ \ L^r(\R^3)\ \ \hbox{as}\ \ n\to\infty,\ \ 1\leq r<3,
\end{equation}
where $\ga:=\sum\limits_{i=1}^2|u_i\ra\la u_i|$. Since $u_i\in\mathcal{H}$ satisfies $(u_i,u_j)=\delta_{ij}$ for $i,j=1,2$, we have
\begin{equation*}
  E_a(2)\leq \E_a(\ga).
\end{equation*}
Moreover, by the weak lower semi-continuity, we obtain from \eqref{2.3} that
\begin{equation*}
  E_a(2)=\lim\limits_{n\to\infty}\E_a(\ga_n)\geq\E_a(\ga),
\end{equation*}
which implies that $\ga$ is a minimizer of $E_a(2)$ for $0< a<a_2^*$.
We then conclude that for any  $0<a<a_2^*$, there exists at least one minimizer of $E_a(2)$.

For any $0<a<a_2^*$, assume  $\ga$ is a minimizer of $E_a(2)$. Similar to the argument of \cite[Appendix A]{Sol}, $\ga$ can be written in the form $\ga=\sum_{i=1}^2|u_{k_i}\ra\la u_{k_i}|$, where $u_{k_i}$ is an eigenfunction of the operator $$H_V=-\Delta+V(x)-\f{5a}{3}\rho_\ga^{\f23}(x)\ \ \mbox{in} \ \ \R^3,$$
and corresponds to the $k_i$-th eigenvalue $\mu_{k_i}$. This gives that $u_{k_i}$ satisfies
\begin{equation}\label{2-2}
  -\Delta u_{k_i}+V(x)u_{k_i}-\f{5a}3\rho_{\ga}^{\f23}u_{k_i}=\mu_{k_i}u_{k_i},\ \ i=1,2,
\end{equation}
where $\rho_\ga(x)=\sum_{j=1}^2|u_{k_j}|^2$.
In the following, we prove that $(u_{k_1},u_{k_2})$ is a ground state of \eqref{0.11}.
Noting from \cite[Theorem 11.8]{Lieb} that $\mu_1<\mu_2$, it suffices to show that $\mu_{k_1}$ and $\mu_{k_2}$ are the 2-first eigenvalues of the operator $H_V$, i.e., $\mu_{k_i}=\mu_i$ holds for $i=1,2$.

We first prove that $\mu_{k_1}=\mu_1$.
On the contrary, suppose $\mu_{k_1}\neq\mu_1$, which  then yields that $\mu_1<\mu_{k_1}\leq\mu_{k_2}$. Hence,  there is an eigenfunction $u_1\in\mathcal{H}$ of $H_V$ in $\R^3$, which corresponds to the first eigenvalue $\mu_1$ and satisfies $(u_1,u_{k_2})=\delta_{1k_2}$.
Define the operator
\begin{equation*}
  \ga':=\ga-|u_{k_1}\ra\la u_{k_1}|+|u_1\ra\la u_1|
  =|u_1\ra\la u_1|+|u_{k_2}\ra\la u_{k_2}|.
\end{equation*}
We then calculate from \eqref{2-2} that
\begin{equation*}
\begin{aligned}
  \mathrm{Tr}(-\Delta \ga')=&\mathrm{Tr}(-\Delta\ga)-\inte|\nabla u_{k_1}|^2dx+\inte |\nabla u_1|^2dx\\
  =&\mathrm{Tr}(-\Delta\ga)+\inte V(x)\big(|u_{k_1}|^2-|u_1|^2\big)dx
  +\f{5a}{3}\inte\rho_{\ga}^{\f23}\big(|u_1|^2-|u_{k_1}|^2\big)dx\\
  &+\mu_1-\mu_{k_1},
\end{aligned}
\end{equation*}
and
\begin{equation*}
  \mathrm{Tr}(V(x)\ga')=\mathrm{Tr}(V(x)\ga)+\inte V(x)\big(|u_1|^2-|u_{k_1}|^2\big)dx.
\end{equation*}
Moreover, by the convexity of $t\mapsto t^{\f53}$ we get that
\begin{equation*}
  \inte (\rho_\ga')^{\f53}dx
  =\inte \big(\rho_{\ga}+|u_1|^2-|u_{k_1}|^2\big)^{\f53}dx
  \geq \inte\rho_\ga^{\f53}dx+\f53\inte\rho_\ga^{\f23}\big(|u_1|^2-|u_{k_1}|^2\big)dx.
\end{equation*}
Since $\mu_1<\mu_{k_1}$, we now conclude from above that
\begin{equation*}
\begin{aligned}
  E_a(2)\leq \E_a(\ga')
  \leq \E_a(\ga)+\mu_1-\mu_{k_1}<\E_a(\ga)=E_a(2),
\end{aligned}
\end{equation*}
a contradiction. We hence obtain that $\mu_{k_1}=\mu_1$.

We next prove that $\mu_{k_2}=\mu_2$.
On the contrary, suppose $\mu_{k_2}\neq\mu_2$. We then deduce from above that $\mu_{k_1}=\mu_1<\mu_2<\mu_{k_2}$. Hence, there exists an eigenfunction $u_2\in\mathcal{H}$ of $H_V$ in $\R^3$, which corresponds  to the second eigenvalue $\mu_2$ and satisfies $(u_{k_1},u_2)=\delta_{k_12}$.
By considering the following operator
\begin{equation*}
  \ga':=\ga-|u_{k_2}\ra\la u_{k_2}|+|u_2\ra\la u_2|
  =|u_{k_1}\ra\la u_{k_1}|+|u_2\ra\la u_2|,
\end{equation*}
the similar argument as  above then yields again a contradiction. This proves that $\mu_{k_2}=\mu_2$. We therefore conclude that  $\mu_{k_i}=\mu_i$ holds for $i=1,2$, which implies that $(u_{k_1},u_{k_2})$ is a ground state of \eqref{0.11}.

2. We next prove the non-existence of minimizers for $E_a(2)$ in the case $a\geq a_2^*$. Let $\ga^{(2)}=\sum_{i=1}^2|Q_i\ra\la Q_i|$ be a minimizer  of $a_2^*$, where $Q_i\in H^1(\R^3)$ satisfies $(Q_i,Q_j)=\delta_{ij}$ for $i,j=1,2$. Take a non-negative function $\varphi(x)\in C_0^\infty(\R^3,[0,1])$, such that $\varphi(x)\equiv 1$ for $|x|\leq1$ and $\varphi(x)\equiv0$ for $|x|\geq 2$.
For any $x_0\in\R^3$ and $\tau>0$, define
\begin{equation}\label{2.16}
  Q_i^\tau(x):=A_i^\tau\tau^{\f32}\varphi(x-x_0)Q_i\big(\tau(x-x_0)\big),\ \ i=1,2,\ \ \mbox{and}\ \  \ga_\tau^{(2)}:=\sum\limits_{i=1}^2|Q_i^\tau\ra\la Q_i^\tau|,
\end{equation}
where $A_i^\tau>0$ is chosen such that $\inte |Q_i^\tau|^2dx=1$ for $i=1,2$. By the exponential decay of $|Q_i|$ in Lemma \ref{lemA.1}, we then derive   that
\begin{equation*}
  \f{1}{(A_i^\tau)^2}=\tau^3\inte \varphi^2(x-x_0)Q_i^2\big(\tau(x-x_0)\big)dx=1+O(\tau^{-\infty})\ \ \hbox{as}\ \ \tau\to\infty,\ \ i=1,2.
\end{equation*}
 Here and below we denote $f(t)=O(t^{-\infty})$, if the function $f(t)$ satisfies $\lim_{t\to\infty}|f(t)|t^s=0$ for all $s>0$.
We therefore obtain that
\begin{equation}\label{2.17}
  A_i^\tau=1+O(\tau^{-\infty}),\ \ i=1,2,\ \ \hbox{and}\ \ a_\tau:=(Q_1^\tau,Q_2^\tau)=O(\tau^{-\infty})\ \ \hbox{as}\ \ \tau\to\infty,
\end{equation}
where we have also used the fact  $(Q_i,Q_j)=\delta_{ij}$ for $i,j=1,2$.
It follows from \eqref{2.17} that the Gram matrix
\begin{equation}\label{2.18}
  G_\tau:=\left(
            \begin{array}{c}
              Q_1^\tau \\
              Q_2^\tau \\
            \end{array}
          \right)
          \left(
            \begin{array}{cc}
              Q_1^\tau & Q_2^\tau \\
            \end{array}
          \right)
          =\left(
             \begin{array}{cc}
               1 & (Q_1^\tau,Q_2^\tau) \\
               (Q_2^\tau,Q_1^\tau) & 1 \\
             \end{array}
           \right)
           =\left(
             \begin{array}{cc}
               1 & a_\tau \\
               a_\tau & 1 \\
             \end{array}
           \right)
\end{equation}
is positive definite for $\tau>0$ large enough.

For $\tau>0$ large enough,  defining
\begin{equation}\label{2.19}
\left(
  \begin{array}{cc}
    \widetilde Q_1^\tau & \widetilde Q_2^\tau \\
  \end{array}
\right)
:=\left(
  \begin{array}{cc}
    Q_1^\tau & Q_2^\tau \\
  \end{array}
\right)
G_\tau^{-\f12},
\end{equation}
it then follows  from \eqref{2.18} that
\begin{equation*}
  \big(\widetilde Q_i^\tau,\widetilde Q_j^\tau\big)=\delta_{ij},\ \ i,j=1,2.
\end{equation*}
Moreover, using  Taylor's expansion, one can obtain from \eqref{2.18} that
\begin{equation*}
  G_\tau^{-\f12}=I_2-\f12a_\tau\left(
             \begin{array}{cc}
               0 & 1 \\
               1 & 0 \\
             \end{array}
           \right)+O(a_\tau^2)\ \ \hbox{as}\ \ \tau\to\infty,
\end{equation*}
 where $I_2$ denotes the 2-order identity matrix. We hence deduce from \eqref{2.19} that
\begin{equation}\label{2.21}
  \left(
  \begin{array}{cc}
    \widetilde Q_1^\tau & \widetilde Q_2^\tau \\
  \end{array}
\right)=
\left(
  \begin{array}{cc}
    Q_1^\tau & Q_2^\tau \\
  \end{array}
\right)-\f12 a_\tau\left(
  \begin{array}{cc}
    Q_2^\tau & Q_1^\tau \\
  \end{array}
\right)+ O(a_\tau^{2}) \ \ \hbox{as}\ \ \tau\to\infty.
\end{equation}
Following  Lemma \ref{lemA.1}, one can  derive from  \eqref{2.16}, \eqref{2.17} and \eqref{2.21} that for $\tau>0$ large enough,
\begin{equation*}
\begin{aligned}
  \inte V(x)|\widetilde Q_i^\tau|^2dx&=
  \inte V(x)\Big[Q_i^\tau-\f12 a_\tau Q_j^\tau+O(a_\tau^2)\Big]^2dx\\
  &=\inte V\Big(\f{x}{\tau}+x_0\Big)\varphi^2\Big(\f{x}{\tau}\Big)Q_i^2(x)dx+O(\tau^{-\infty})\\
  &\leq\int_{|x|\leq 2\tau}V\l(\f{x}{\tau}+x_0\r)Q_i^2(x)dx+O(\tau^{-\infty})\\
  &\leq C\inte Q_i^2(x)dx+O(\tau^{-\infty})<\infty,\ \ i=1,2,
\end{aligned}
\end{equation*}
and similarly,
\begin{equation*}
  \inte|\nabla \widetilde Q_i^\tau|^2dx=\tau^2\inte|\nabla Q_i|^2dx+O(\tau^{-\infty})<\infty,\ \ i=1,2,
\end{equation*}
due to the fact that $Q_i\in H^1(\R^3)$ holds for $i=1,2$.
This  implies that for $\tau>0$ large enough, $\widetilde Q_i^\tau\in \mathcal{H}$ holds for $i=1,2$.

For $\tau>0$ large enough, denoting
 \begin{equation}\label{2.27}
   \widetilde\ga_\tau^{(2)}:=\sum\limits_{i=1}^2|\widetilde Q_i^\tau\ra\la \widetilde Q_i^\tau|,
\end{equation}
where $\widetilde Q_1^\tau$ and $\widetilde Q_2^\tau$ are as in \eqref{2.19},
 we next estimate each term of $\E_a(\widetilde\ga_\tau^{(2)})$.
 It follows from Lemma \ref{lemA.1}, \eqref{2.16}, \eqref{2.17} and \eqref{2.21} that
 \begin{equation}\label{2.23}
 \begin{aligned}
   &\mathrm{Tr}(-\Delta\widetilde\ga_\tau^{(2)})-a\inte \rho_{\widetilde\ga_\tau^{(2)}}^{\f53}dx\\
   =&\sum\limits_{i=1}^2\inte |\nabla \widetilde Q_i^\tau|^2dx-a\inte\Big(\sum\limits_{j=1}^2|\widetilde Q_j^\tau|^2\Big)^{\f53}dx\\
   =&\tau^2\Big[\mathrm{Tr}(-\Delta\ga^{(2)})-a\inte\rho_{\ga^{(2)}}^{\f53}dx\Big]
   +O(\tau^{-\infty})\\
   =&(a_2^*-a)\tau^2\inte\rho_{\ga^{(2)}}^{\f53}dx+O(\tau^{-\infty})\ \ \hbox{as}\ \ \tau\to\infty,
 \end{aligned}
 \end{equation}
due to the fact that $\ga^{(2)}$ is a minimizer of $a_2^*$ with $\|\ga^{(2)}\|=1$.
On the other hand, since the function $x\mapsto V(x)\varphi^2(x-x_0)$ is bounded and has compact support,  we deduce from Lemma \ref{lemA.1}, \eqref{2.16}, \eqref{2.17} and \eqref{2.21}  that
 \begin{equation}\label{2.24}
 \begin{aligned}
   \lim\limits_{\tau\to\infty}\mathrm{Tr}\big(V(x)\widetilde\ga_\tau^{(2)}\big)
   &=\lim\limits_{\tau\to\infty}\sum\limits_{i=1}^2\inte V(x)|\widetilde Q_i^\tau|^2dx\\
   &=\lim\limits_{\tau\to\infty}\sum\limits_{i=1}^2\inte V\l(\f x\tau+x_0\r)\varphi^2\l(\f x\tau\r)Q_i^2(x)dx\\
   &=V(x_0)\inte\rho_{\ga^{(2)}}(x)dx=2V(x_0).
 \end{aligned}
 \end{equation}
 Combining \eqref{2.23} with \eqref{2.24} yields that for $a> a_2^*$,
 \begin{equation*}
 \begin{aligned}
   E_a(2)\leq& \lim\limits_{\tau\to\infty} \E_a(\widetilde\ga_\tau^{(2)})\\
   =&\lim\limits_{\tau\to\infty}\Big\{\mathrm{Tr}(-\Delta+V(x))\widetilde\ga_\tau^{(2)}-a\inte \rho_{\widetilde\ga_\tau^{(2)}}^{\f53}dx\Big\}
   =-\infty,
 \end{aligned}
 \end{equation*}
 and hence there is no  minimizer of $E_a(2)$ for $a>a_2^*$.

As for the case $a=a_2^*$, taking the infimum over $x_0\in\R^3$,
  it then follows from \eqref{2.1}, \eqref{2.23} and \eqref{2.24} that $E_{a_2^*}(2)=0$. We next prove the non-existence of minimizers for $E_{a_2^*}(2)$. On the contrary, assume that $\ga=\sum_{i=1}^2|u_i\ra\la u_i|$, where $u_i\in\mathcal{H}$ and  $(u_i,u_j)=\delta_{ij}$ for $i,j=1,2$, is a minimizer of $E_{a_2^*}(2)$. We then obtain from \eqref{0.7} that for $V(x)\geq 0$,
 \begin{equation}\label{2.25}
   \inte V(x)\rho_{\ga}(x)dx=0,
 \end{equation}
 and
 \begin{equation}\label{2.26}
   \mathrm{Tr}(-\Delta\ga)=a_2^*\inte\rho_{\ga}^{\f53}(x)dx.
 \end{equation}
Since $\lim_{|x|\to\infty}V(x)=\infty$, we derive from \eqref{2.25} that  $\rho_\ga(x)$ has compact support. Following \eqref{2.26}, one can obtain that $\ga$ is a minimizer of $a_2^*$, which implies from \cite[Theorem 6]{Frank2}  that $u_1(x)$ and $u_2(x)$ are the 2-first eigenfunctions of the operator $-\Delta-\f53a_2^*\rho_\ga^{2/3}(x)$ in $\R^3$.  Hence $\rho_\ga(x)=u_1^2(x)+u_2^2(x)>0$ in $\R^3$, in view of the fact (cf. \cite[Theorem 11.8]{Lieb}) that the first eigenfunction $u_1(x)$ satisfies $u_1^2(x)>0$ in $\R^3$. This is however a contradiction, and therefore there is no minimizer for $E_{a_2^*}(2)$. This completes the proof of Theorem \ref{thm1}.\qed

Note from the proof of Theorem \ref{thm1} that $\lim_{a\nearrow a_2^*}E_{a}(2)=E_{a_2^*}(2)=\inf_{x\in\R^3}V(x)=0$. Indeed, by  taking $a\nearrow a_2^*$ and setting $\tau\to\infty$, we derive from \eqref{2.23} and \eqref{2.24} that $\limsup\limits_{a\nearrow a_2^*}E_a(2)\leq 2V(x_0)$. The above result can be then obtained by taking the infimum over $x_0\in\R^3$.

\section{Estimates  of Minimizers  as $a\nearrow a_2^*$ }
Assume $V(x)$ satisfies \eqref{0-2}, it follows from Theorem \ref{thm1} that $E_a(2)$ admits minimizers, if and only if $0<a<a_2^*$.  In this section, we shall  establish some refined estimates of minimizers for $E_a(2)$ as $a\nearrow a_2^*$. We first address the following energy estimates of $E_a(2)$ as $a\nearrow a_2^*$.

\begin{lem}\label{lem3.1}
Assume $V(x)$ satisfies \eqref{0-2}. Then there exist two positive constants $m$ and $M$, independent of $0<a<a_2^*$, such that
\begin{equation}\label{3.5}
  0<m(a_2^*-a)^{\f{p}{p+2}}\leq E_a(2)\leq M (a_2^*-a)^{\f{p}{p+2}}\ \ \hbox{as}\ \ a\nearrow a_2^*,
\end{equation}
where $p>0$ is as in \eqref{0-4}.
\end{lem}
\ni{\bf Proof.}
For any $0<a<a_2^*$, $\beta>0$, and $\ga=\sum_{i=1}^2|u_i\ra\la u_i|$, where $u_i\in\mathcal{H}$ and $(u_i,u_j)=\delta_{ij}$ for $i,j=1,2$, we obtain from Young's inequality and \eqref{0.7} that
\begin{equation}\label{3.2}
\begin{aligned}
  \E_a(\ga)\geq& \inte V(x)\rho_{\ga}(x)dx+(a_2^*-a)\inte\rho_\ga^{\f53}(x)dx\\
  =&2\beta+\inte \l(V(x)-\beta\r)\rho_\ga(x) dx+(a_2^*-a)\inte\rho_\ga^{\f53}(x)dx\\
  \geq&2\beta-\inte \big[\beta-V(x)\big]_+\rho_\ga(x)dx+(a_2^*-a)\inte\rho_\ga^{\f53}(x)dx\\
  \geq&2\beta-\f25\l(\f35\r)^{\f32}\f{1}{(a_2^*-a)^{\f32}}\inte \big[\beta-V(x)\big]_+^{\f52}dx,
\end{aligned}
\end{equation}
where $[\cdot]_+=\max\{0,\cdot\}$ denotes the positive part.

For $\beta>0$ small enough, since $V(x)$ satisfies \eqref{0-2},  the set  $\{x\in\R^3: V(x)\leq\beta\}$ is contained in the  union of $l$ disjoint balls, each of which has the center at the minimum point $x_m$ ($m=1,\cdots,l$), together with the  radius no more than $K\beta^{\f1p}$ for some suitable constant $K>0$. Moreover, $V(x)\geq \big(\f{|x-x_m|}{K}\big)^p$ holds in these disjoint balls. We therefore derive that
\begin{equation}\label{3.3}
\begin{aligned}
  \inte\big[\beta-V(x)\big]_+^{\f52}dx
  &\leq l\int_{|x|\leq K\beta^{\f1p}} \Big[\beta-\Big(\f{|x|}{K}\Big)^p\Big]^{\f52}dx\\
  &=lK^3\beta^{\f{5p+6}{2p}}\int_{|x|\leq 1}(1-|x|^p)^{\f52}dx
  \leq\f{4\pi lK^3}{3}\beta^{\f{5p+6}{2p}}.
\end{aligned}
\end{equation}
Applying \eqref{3.2} and \eqref{3.3}, there exists a constant $m>0$ such that
\begin{equation*}
  \E_a(\ga)\geq 2\beta-C_0\f{\beta^{\f{5p+6}{2p}}}{(a_2^*-a)^{\f32}}\geq m(a_2^*-a)^{\f{p}{p+2}}>0,
\end{equation*}
where $C_0:=\f{8\pi l K^3}{15}(\f35)^{\f32}>0$, and the second inequality is derived by taking $\beta=(a_2^*-a)^{\f{p}{p+2}}\l[\f{4p}{(5p+6)C_0}\r]^{\f{2p}{3p+6}}>0$. This gives the lower bound of \eqref{3.5} as $a\nearrow a_2^*$.

In order to derive the upper bound of \eqref{3.5}, we take the test function $\widetilde\ga^{(2)}_\tau$ of the form \eqref{2.27}, where the point $x_0$ in \eqref{2.16} is chosen such that $x_0\in\mathcal{Z}$ defined in \eqref{0-3}. Choose sufficiently small $\mathcal{R}>0$ so that
\begin{equation*}
  V(x)\leq C_1|x-x_0|^p\ \ \hbox{for}\ \ |x-x_0|\leq\mathcal{R}.
\end{equation*}
We therefore obtain from Lemma \ref{lemA.1}, \eqref{2.16}, \eqref{2.17} and \eqref{2.21} that
\begin{equation*}
\begin{aligned}
  \mathrm{Tr}(V\widetilde\ga^{(2)}_\tau)
  &=\sum\limits_{i=1}^2\inte V(x)|\widetilde Q_i^\tau(x)|^2dx\\
  &=\sum\limits_{i=1}^2\inte V\l(\f{x}{\tau}+x_0\r)\varphi^2\l(\f{x}{\tau}\r)Q_i^2(x)dx+O(\tau^{-\infty})\\
  &\leq C_1\tau^{-p}\inte|x|^p\rho_{\ga^{(2)}}(x) dx+O(\tau^{-\infty})\ \ \hbox{as}\ \ \tau\to\infty,
\end{aligned}
\end{equation*}
which then yields from \eqref{2.23} that
\begin{equation*}
\begin{aligned}
  E_a(2)&\leq \E_a(\widetilde\ga_\tau^{(2)})
  \leq (a_2^*-a)\tau^2\inte\rho_{\ga^{(2)}}^{\f53}(x)dx\\
  &\qquad\qquad\qquad+C_1\tau^{-p}\inte|x|^p\rho_{\ga^{(2)}}(x) dx+O(\tau^{-\infty})\ \ \hbox{as}\ \ \tau\to\infty.
\end{aligned}
\end{equation*}
Setting $\tau=(a_2^*-a)^{-\f{1}{p+2}}>0$ into the above estimate thus gives the upper bound of \eqref{3.5} as $a\nearrow a_2^*$. This therefore completes the proof of Lemma \ref{lem3.1}.\qed

Applying the energy estimates of Lemma \ref{lem3.1}, we next address the following estimates of $\rho_{\ga_a}(x)$ as $a\nearrow a_2^*$, where $\ga_a$ is a minimizer of $E_a(2)$.

\begin{lem}\label{lem3.2}
Assume $V(x)$ satisfies \eqref{0-2}, and suppose $\ga_a=\sum_{i=1}^2|u_i^a\ra\la u_i^a|$ is a minimizer of $E_a(2)$, where $u_i^a\in\mathcal{H}$ satisfies  $(u_i^a,u_j^a)=\delta_{ij}$ for $i,j=1,2$. Then there exists a constant $L>0$, independent of $0<a<a_2^*$, such that
\begin{equation}\label{3.6}
  0<L(a_2^*-a)^{-\f{2}{p+2}}\leq \inte\rho_{\ga_a}^{\f53}(x)dx\leq \f1L(a_2^*-a)^{-\f{2}{p+2}}\ \ \hbox{as}\ \ a\nearrow a_2^*,
\end{equation}
where $p>0$ is as in \eqref{0-4}, and  $\rho_{\ga_a}(x)=\sum_{i=1}^2|u_i^a(x)|^2$.
\end{lem}

\ni{\bf Proof.}
By Lemma \ref{lem3.1}, it follows from \eqref{0.7} and \eqref{0-2} that
\begin{equation*}
  M(a_2^*-a)^{\f{p}{p+2}}\geq E_a(2)\geq (a_2^*-a)\inte\rho_{\ga_a}^{\f53}(x)dx\ \ \hbox{as}\ \ a\nearrow a_2^*,
\end{equation*}
which yields the upper bound of \eqref{3.6}.

We next prove the lower bound of \eqref{3.6}. For any $0<b<a<a_2^*$, we derive that
\begin{equation*}
  E_b(2)\leq\E_b(\ga_a)=E_a(2)+(a-b)\inte\rho_{\ga_a}^{\f53}(x)dx.
\end{equation*}
Following Lemma \ref{lem3.1}, we have
\begin{equation*}
\begin{aligned}
  \inte\rho_{\ga_a}^{\f53}(x)dx&\geq \f{E_b(2)-E_a(2)}{a-b}\geq\f{m(a_2^*-b)^{\f{p}{p+2}}
  -M(a_2^*-a)^{\f{p}{p+2}}}{a-b}\\
  &=(a_2^*-a)^{-\f{2}{p+2}}\f{m(1+\delta)^{\f{p}{p+2}}-M}{\delta}\ \ \hbox{as}\ \ a\nearrow a_2^*,
\end{aligned}
\end{equation*}
by taking $b=a-\delta(a_2^*-a)\in (0,a)$. When $a>0$ is sufficiently close to $a_2^*$,  one can choose sufficiently large $\delta>0$, so that  the last  fraction of the above estimate is positive. This gives the lower bound of \eqref{3.6}, and the proof of Lemma \ref{lem3.2} is therefore complete. \qed

Under the assumption  \eqref{0-2}, we now define
\begin{equation}\label{3.7}
  \vaa:=(a_2^*-a)^{\f{1}{p+2}}>0,\ \ 0<a<a_2^*,
\end{equation}
where $p>0$ is as in \eqref{0-4}. The following lemma is then concerned with the analysis properties of minimizers for  $E_a(2)$ in terms of $\vaa >0$.

\begin{lem}\label{lem3.3}
Assume $V(x)$ satisfies \eqref{0-2}, and suppose $\ga_a=\sum_{i=1}^2|u_i^a\ra\la u_i^a|$ is a minimizer of $E_a(2)$, where $u_i^a\in\mathcal{H}$ satisfies  \eqref{0.11} and  $(u_i^a,u_j^a)=\delta_{ij}$ for $i,j=1,2$. Then we have
\begin{enumerate}
  \item There exist a sequence $\{y_{\vaa}\}\subset\R^3$, positive constants $R_0$ and $\eta$ such that the sequence
      \begin{equation}\label{3.38}
        \bar w_i^a(x):=\vaa^{\f32}u_i^a( \vaa x+\vaa y_{\vaa}),\ \ i=1,2,\ \ \bar\ga_a:=\sum\limits_{i=1}^2|\bar w_i^a\ra\la\bar w_i^a|,
      \end{equation}
      satisfies
      \begin{equation}\label{3.8}
        \liminf_{a\nearrow a_2^*}\int_{B_{R_0}(0)}\rho_{\bar\ga_a}(x)dx\geq \eta>0,
      \end{equation}
      where $\rho_{\bar\ga_a}(x):=\sum_{i=1}^2\l|\bar w_i^a(x)\r|^2$, and $\vaa>0$ is defined by \eqref{3.7}.
  \item The  point $\bar x_a:=\vaa y_{\vaa}$ satisfies
  \begin{equation}\label{03-1}
    \lim\limits_{a\nearrow a_2^*}\mathrm{dist}(\bar x_a,\Lambda)=0,
  \end{equation}
  where the set $\Lambda$ is defined by \eqref{0-4}.
  Moreover, for any sequence $\{a_n\}$ satisfying $a_n\nearrow a_2^*$ as $n\to\infty$, there exist a subsequence, still denoted by $\{a_n\}$, of $\{a_n\}$ and a point $x_k\in\Lambda$ such that
  \begin{equation}\label{3-1}
    \bar x_{a_n}\stackrel{n}{\longrightarrow}
 x_k\ \ \hbox{and}\ \ \bar w_i^{a_n}(x):=\va_{a_n}^{\f32}u_i^{a_n}(\va_{a_n} x+\bar x_{a_n})\stackrel{n}{\longrightarrow} \bar w_i(x)
  \end{equation}
  strongly in $H^1(\R^3)$, where
   $\bar\ga:=\sum_{i=1}^2|\bar w_i\ra\la \bar w_i|$ is a minimizer of $a_2^*$ defined by \eqref{0.7}.
\end{enumerate}
\end{lem}
\ni{\bf Proof.} 1. Assume $\ga_a=\sum_{i=1}^2|u_i^a\ra\la u_i^a|$ is a minimizer of $E_a(2)$, where $u_i^a\in\mathcal{H}$ satisfies $(u_i^a,u_j^a)=\delta_{ij}$ for $i,j=1,2$. Applying Lemma \ref{lem3.1}, it then follows from \eqref{0.7} and \eqref{0-2} that
\begin{equation*}
  0\leq \mathrm{Tr}(-\Delta\ga_a)-a\inte \rho_{\ga_a}^{\f53}(x)dx\leq E_a(2)\to0\ \ \hbox{as}\ \ a\nearrow a_2^*.
\end{equation*}
Note from Lemma \ref{lem3.2} that $\lim\limits_{a\nearrow a_2^*}\inte\rho_{\ga_a}^{\f53}(x)dx\to\infty$, and hence
\begin{equation*}
  0\leq\f{\mathrm{Tr}(-\Delta\ga_a)}{\inte\rho_{\ga_a}^{\f53}(x)dx}
  -a\leq\f{E_a(2)}{\inte\rho_{\ga_a}^{\f53}(x)dx}\to0\ \ \hbox{as}\ \ a\nearrow a_2^*,
\end{equation*}
which gives that
\begin{equation*}
  \f{\mathrm{Tr}(-\Delta\ga_a)}{\inte\rho_{\ga_a}^{\f53}(x)dx}\to a_2^*\ \ \hbox{as}\ \ a\nearrow a_2^*.
\end{equation*}
Taking $m_1=\max\{\f{3a_2^*}{2},\f{2}{a_2^*}\}$, it yields that
\begin{equation*}
  0<\f1{m_1}\inte\rho_{\ga_a}^{\f53}(x)dx\leq \mathrm{Tr}\big(-\Delta\ga_a\big)\leq m_1\inte\rho_{\ga_a}^{\f53}(x)dx\ \ \hbox{as}\ \ a\nearrow a_2^*.
\end{equation*}
We then deduce from Lemma \ref{lem3.2} that there exists $C_2:=\f{m_1}{L}>0$  such that
\begin{equation}\label{3.10}
  0<\f{1}{C_2}(a_2^*-a)^{-\f{2}{p+2}}\leq \mathrm{Tr}(-\Delta\ga_a)\leq C_2(a_2^*-a)^{-\f{2}{p+2}}\ \ \hbox{as}\ \ a\nearrow a_2^*.
\end{equation}

 Denote
\begin{equation*}
  \widetilde w_i^a(x):=\vaa^{\f32} u_i^a(\vaa x),\ \ i=1,2,\ \ \widetilde\ga_{a}:=\sum\limits_{i=1}^2|\widetilde w_i^a\ra\la \widetilde w_i^a|,
\end{equation*}
where $\vaa>0$ is as in \eqref{3.7}.
 It then follows from Lemma \ref{lem3.2} and \eqref{3.10} that
\begin{equation}\label{3.11}
  0<\f{1}{C_2}\leq \mathrm{Tr}(-\Delta\widetilde\ga_a)\leq C_2\ \ \hbox{and}\ \  0< L\leq\inte\rho_{\widetilde\ga_a}^{\f53}(x)dx\leq  \f1 L\ \ \hbox{as}\ \ a\nearrow a_2^*.
\end{equation}
On the other hand, the Hoffmann-Ostenhof  inequality \cite{Hof} gives that
\begin{equation}\label{3.53}
  \mathrm{Tr}(-\Delta \widetilde\ga_a)\geq \inte |\nabla\sqrt{\rho_{\widetilde\ga_a}}|^2dx.
\end{equation}
 We therefore deduce from \eqref{3.11} and \eqref{3.53} that the sequence $\{\sqrt{\rho_{\widetilde\ga_a}}\}$ is bounded uniformly in $H^1(\R^3)$ as $a\nearrow a_2^*$.

We next claim that there exist a sequence $\{y_{\vaa} \}\subset\R^3$,  $R_0>0$ and $\eta>0$ such that
\begin{equation}\label{3.12}
  \liminf_{a\nearrow a_2^*}\int_{B_{R_0}(y_{\vaa})}\rho_{\widetilde\ga_a}(x)dx\geq \eta>0.
\end{equation}
Indeed, if \eqref{3.12} is not true, then for any $R>0$, there exists a sequence $\{a_n\}$, where $a_n\nearrow a_2^*$ as $n\to\infty$, such that
\begin{equation*}
  \lim\limits_{n\to\infty}\sup_{y\in\R^3}\int_{B_R(y)}\rho_{\widetilde\ga_{a_n}}(x) dx=0.
\end{equation*}
Since the sequence $\{\sqrt{\rho_{\widetilde\ga_{a_n}}}\}$ is bounded uniformly in $H^1(\R^3)$ as $n\to\infty$, we derive from \cite[Theorem 1.21]{Willem} that $\rho_{\widetilde\ga_{a_n}}(x)\to0$ strongly in $L^q(\R^3)$ as $n\to\infty$ for $1<q<3$. This is however a contradiction in view of \eqref{3.11}. We therefore obtain that the claim \eqref{3.12} holds true, which further yields that \eqref{3.8} holds true.

2. We first prove that \eqref{03-1} holds true.
On the contrary, assume that  \eqref{03-1} is not true. Then there exist   a sequence $\{a_n\}$, where $a_n\nearrow a_2^*$ as $n\to\infty$, and a constant $\delta>0$ such that
\begin{equation*}
    \mathrm{dist}(\bar x_{a_n},\Lambda)\geq\delta>0\ \ \hbox{as}\ \ n\to\infty,
  \end{equation*}
  which yields that there exists a constant $C(\delta)>0$ such that
\begin{equation*}
  V(\bar x_{a_n})\geq C(\delta)>0\ \ \hbox{as}\ \ n\to\infty.
\end{equation*}
By Fatou's lemma, we therefore derive from \eqref{3.8} that
\begin{equation}\label{3.13}
\begin{aligned}
  \liminf_{n\to\infty}\inte V(\va_{a_n} x+\bar x_{a_n})\rho_{\bar\ga_{a_n}}(x)dx
  &\geq\int_{B_{R_0}(0)}\liminf_{n\to\infty}V(\va_{a_n} x+\bar x_{a_n})\rho_{\bar\ga_{a_n}}(x)dx\\
  &\geq \f{C(\delta)}{2}\eta>0.
\end{aligned}
\end{equation}
On the other hand, one can deduce from \eqref{0.7} and Lemma \ref{lem3.1} that
\begin{equation}\label{3-3}
  0\leq\inte V(\va_{a_n} x+\bar x_{a_n})\rho_{\bar\ga_{a_n}}(x)dx=\inte V(x)\rho_{\ga_{a_n}}(x)dx\leq E_{a_n}(2)\to0\ \ \hbox{as}\ \ n\to\infty,
\end{equation}
which however contradicts with \eqref{3.13}, and hence  \eqref{03-1} holds true.

We now focus on the proof of \eqref{3-1}. Towards this purpose, we first claim that
\begin{equation}\label{3.25}
  \mathrm{Tr}(-\Delta\bar\ga_a)=a_2^*\inte\rho_{\bar\ga_a}^{\f53}(x)dx+o(1)\ \ \hbox{as}\ \ a\nearrow a_2^*,
\end{equation}
where $\bar\ga_a$ is defined by \eqref{3.38}.
Indeed, note that
$\ga_a=\sum_{i=1}^2|u_i^a\ra\la u_i^a|$ is a minimizer of $E_a(2)$, where $(u_1^a,u_2^a)$ satisfies the following system 
\begin{equation}\label{3.15}
  -\Delta u_i^a+V(x)u_i^a-\f{5a}{3}\rho_{\ga_a}^{\f23}u_i^a=\mu_i^au_i^a\ \ \hbox{in}\ \ \R^3,\ \ i=1,2.
\end{equation}
Here $\rho_{\ga_a}(x)=\sum_{i=1}^2|u_i^a(x)|^2$, and $\mu_1^a<\mu_2^a$ are the 2-first eigenvalues of the operator $-\Delta+V(x)-\f{5a}{3}\rho_{\ga_a}^{\f23}$ in $\R^3$.
We hence deduce from Lemma \ref{lem3.1} and \eqref{3.15}  that
\begin{equation}\label{3.24}
  \sum\limits_{i=1}^2\mu_i^a\vaa^2
  =\vaa^2E_a(2)-\f{2a}{3}\vaa^2\inte\rho_{\ga_a}^{\f53}dx
  =-\f{2a}{3}\inte\rho_{\bar\ga_a}^{\f53}(x)dx+o(1)\ \ \hbox{as} \ \ a\nearrow a_2^*.
\end{equation}


On the other hand, we obtain from \eqref{3.38} and   \eqref{3.15}  that
\begin{equation}\label{3.16}
  -\Delta \bar w_i^a+\vaa^2 V(\vaa x+\bar x_a)\bar w_i^a-\f{5a}{3}\rho_{\bar\ga_a}^{\f23}\bar w_i^a=\mu_i^a\vaa^2\bar w_i^a\ \ \hbox{in}\ \ \R^3,\ \ i=1,2,
\end{equation}
which implies that
\begin{equation}\label{3.22}
  \mathrm{Tr}(-\Delta\bar\ga_a)+\vaa^2\inte V(\vaa x+\bar x_a)\rho_{\bar\ga_a}(x)dx-\f{5a}{3}\inte\rho_{\bar\ga_a}^{\f53}(x)dx
  =\sum\limits_{i=1}^2\mu_i^a\vaa^2.
\end{equation}
It follows from \eqref{3-3} that
\begin{equation*}
  \vaa^2\inte V(\vaa x+\bar x_a)\rho_{\bar\ga_a}(x)dx\to0\ \ \hbox{as}\ \ a\nearrow a_2^*,
\end{equation*}
which and \eqref{3.22} give that
\begin{equation}\label{3.23}
  \mathrm{Tr}(-\Delta\bar\ga_a)-\f{5a}{3}\inte\rho_{\bar\ga_a}^{\f53}(x)dx
  =\sum\limits_{i=1}^2\mu_i^a\vaa^2+o(1)\ \ \hbox{as}\ \ a\nearrow a_2^*.
\end{equation}
Combining \eqref{3.24} with \eqref{3.23} yields that
\begin{equation*}
  \mathrm{Tr}(-\Delta\bar\ga_a)=a_2^*\inte\rho_{\bar\ga_a}^{\f53}(x)dx+o(1)\ \ \hbox{as}\ \ a\nearrow a_2^*,
\end{equation*}
and hence the claim \eqref{3.25} holds true.

Let $\{a_n\}$ be any sequence satisfying $a_n\nearrow a_2^*$ as $n\to\infty$. It follows from \eqref{03-1} that there exist a subsequence, still denoted by $\{a_n\}$, of $\{a_n\}$ and a point $x_k\in\Lambda$ such that
\begin{equation}\label{03-2}
  \bar x_{a_n}\to x_k\ \ \hbox{as}\ \ n\to\infty.
\end{equation}
Similar to \eqref{3.11}, we obtain that  $\{ \bar w_i^{a_n}\}$ is bounded uniformly in $H^1(\R^3)$ as $n\to\infty$ for $i=1,2$. Hence,   up to a subsequence if necessary, there exists a function $\bar w_i\in H^1(\R^3)$ such that
\begin{equation}\label{3.18}
  \bar w_i^{a_n}\rightharpoonup \bar w_i\ \ \hbox{weakly in}\ \ H^1(\R^3)\ \ \ \ \hbox{as}\ \ n\to\infty,\ \ i=1,2,
\end{equation}
and
\begin{equation*}
    \bar w_i^{a_n}\to \bar w_i\ \ \hbox{strongly in}\  \  L^q_{loc}(\R^3)\ \ \hbox{as}\ \ n\to\infty,\ \  2\leq q<6,\ \ i=1,2.
\end{equation*}
This gives that $$\bar w_i^{a_n}\to \bar w_i\ \ \hbox{a.e. in}\ \ \R^3\ \ \hbox{as}\ \ n\to\infty,\ \ i=1,2,$$ and
\begin{equation*}
  \rho_{\bar\ga_n}\to\rho_{\bar\ga}:=\bar w_1^2+\bar w_2^2 \ \ \hbox{strongly in} \ \ L^r_{loc}(\R^3)\ \ \hbox{as}\ \ n\to\infty,\ \  1\leq r<3,
\end{equation*}
where we denote $\bar\ga_n:=\bar\ga_{a_n}$ and $\bar\ga:=\sum_{i=1}^2|\bar w_i\ra\la \bar w_i|$.

By an adaptation of the classical dichotomy result (cf. \cite[Section 3.3]{Mar}), one can deduce from \eqref{3.8} that up to a subsequence of $\{a_n\}$ if necessary,  there exists a sequence $\{R_n\}$ with $R_n\to\infty$ as $n\to \infty$ such that
\begin{equation}\label{3.29}
  0<\lim\limits_{n\to\infty}\int_{|x|\leq R_n}\rho_{\bar\ga_n}dx=\inte\rho_{\bar\ga} dx\ \ \hbox{and}\ \ \lim\limits_{n\to\infty}\int_{R_n\leq|x|\leq 2R_n}\rho_{\bar\ga_n}dx=0.
\end{equation}
Let $\chi(x)\in  C_0^\infty(\R^3,[0,1])$ be a cut-off function satisfying $\chi(x)\equiv1$ for $|x|\leq1$ and $\chi(x)\equiv0$ for $|x|\geq2$. Taking $\chi_n(x):=\chi(\f{x}{R_n})$ and $\eta_n(x)=\sqrt{1-\chi_n^2(x)}$, we then obtain from \eqref{3.29} that
\begin{equation}\label{3.30}
  \chi_n^2\rho_{\bar\ga_n}\to\rho_{\bar\ga}\ \ \hbox{strongly in}\ \ L^r(\R^3)\ \ \hbox{as}\ \ n\to\infty,\ \ 1\leq r<3.
\end{equation}
Following the IMS formula \cite[Theorem 3.2]{Cycon} and Fatou's lemma \cite[Theorem 2.7]{Simon}, we derive   that
\begin{equation}\label{3.32}
\begin{aligned}
  \mathrm{Tr}(-\Delta\bar\ga_n)&=\mathrm{Tr}(-\Delta\chi_n\bar\ga_n\chi_n)
  +\mathrm{Tr}(-\Delta\eta_n\bar\ga_n\eta_n)
  -\inte(|\nabla\chi_n|^2+|\nabla\eta_n|^2)\rho_{\bar\ga_n}dx\\
  &\geq\mathrm{Tr}(-\Delta\chi_n\bar\ga_n\chi_n)
  +\mathrm{Tr}(-\Delta\eta_n\bar\ga_n\eta_n)
  -2CR_n^{-2}\\
  &=\mathrm{Tr}(-\Delta\chi_n\bar\ga_n\chi_n)
  +\mathrm{Tr}(-\Delta\eta_n\bar\ga_n\eta_n)+o(1)\\
  &\geq \mathrm{Tr}(-\Delta\bar \ga)+\mathrm{Tr}(-\Delta\eta_n\bar\ga_n\eta_n)+o(1)\ \ \hbox{as}\ \ n\to\infty.\\
\end{aligned}
\end{equation}
Moreover, we deduce from \eqref{3.30} that
\begin{equation}\label{3.33}
\begin{aligned}
  \inte\rho_{\bar\ga_n}^{\f53}dx
  &=\inte\chi_n^2\rho_{\bar\ga_n}^{\f53}dx
  +\inte(\eta_n^2\rho_{\bar\ga_n})^{\f53}dx
  +\inte (\eta_n^2-\eta_n^{\f{10}{3}})\rho_{\bar\ga_n}^{\f53}dx\\
  &=\inte\rho_{\bar\ga}^{\f53}dx+\inte(\eta_n^2\rho_{\bar\ga_n})^{\f53}dx+o(1)\ \ \hbox{as}\ \ n\to\infty.
\end{aligned}
\end{equation}

Since $\|\bar\ga\|\leq\liminf_{n\to\infty}\|\bar\ga_n\|=1$ and $\|\eta_n\bar\ga_n\eta_n\|\leq \|\bar\ga_n\|=1$, we obtain from \eqref{0.7}, \eqref{3.25}, \eqref{3.32} and \eqref{3.33} that
\begin{equation}\label{3.28}
\begin{aligned}
  0=&\lim\limits_{n\to\infty}\Big\{\mathrm{Tr}(-\Delta\bar\ga_n)
  -a_2^*\inte\rho_{\bar\ga_n}^{\f53}dx\Big\}\\
  \geq&\mathrm{Tr}(-\Delta\bar\ga)-a_2^*\inte\rho_{\bar\ga}^{\f53}dx
  +\lim\limits_{n\to\infty}\Big\{\mathrm{Tr}(-\Delta\eta_n\bar\ga_n\eta_n)
  -a_2^*\inte(\eta_n^2\rho_{\bar\ga_n})^{\f53}dx\Big\}\\
  \geq&\|\bar\ga\|^{\f23}\mathrm{Tr}(-\Delta\bar\ga)-a_2^*\inte\rho_{\bar\ga}^{\f53}dx\\
  &
  +\lim\limits_{n\to\infty}\Big\{\|\eta_n\bar\ga_n\eta_n\|^{\f23}\mathrm{Tr}(-\Delta\eta_n\bar\ga_n\eta_n)
  -a_2^*\inte(\eta_n^2\rho_{\bar\ga_n})^{\f53}dx\Big\}\\
  \geq& \|\bar\ga\|^{\f23}\mathrm{Tr}(-\Delta\bar\ga)-a_2^*\inte\rho_{
  \bar\ga}^{\f53}dx\geq0,
\end{aligned}
\end{equation}
which implies that $\bar\ga$ is a minimizer of $a_2^*$ and $\|\bar\ga\|=1$. It also follows from \cite[Theorem 6]{Frank2} that any minimizer $\ga^{(2)}$ of $a_2^*$ is of the form
\begin{equation*}
  \ga^{(2)}=\|\ga^{(2)}\|\sum\limits_{i=1}^2|Q_i\ra\la Q_i|,\ \ (Q_i,Q_j)=\delta_{ij},\ \ i,j=1,2.
\end{equation*}
We therefore obtain that $\bar\ga=\|\bar\ga\|\sum\limits_{i=1}^2|Q_i\ra\la Q_i|=\sum\limits_{i=1}^2|Q_i\ra\la Q_i|$, and hence
\begin{equation}\label{3.37}
  2=\lim\limits_{n\to\infty}\inte\rho_{\bar\ga_n}dx=\inte\rho_{\bar\ga} dx.
\end{equation}
 Moreover, one can also derive from \eqref{3.28} and \eqref{3.37} that
\begin{equation}\label{3.34}
  \lim\limits_{n\to\infty}\mathrm{Tr}(-\Delta\bar\ga_n)=\mathrm{Tr}(-\Delta\bar\ga)\ \ \hbox{and}\ \ \lim\limits_{n\to\infty}\inte\rho_{\bar\ga_n}^{\f53}dx=\inte\rho_{\bar\ga}^{\f53}dx.
\end{equation}
We then derive from  \eqref{3.37} and \eqref{3.34} that up to a subsequence if necessary,
\begin{equation}\label{3.35}
  \bar w_i^{a_n}(x):=\va_{a_n}^{\f32}u_i^{a_n}(\va_{a_n} x+\bar x_{a_n})\to \bar w_i(x)\ \ \hbox{strongly in}\ \ H^1(\R^3)\ \ \hbox{as}\ \ n\to\infty,
\end{equation}
where $\bar\ga=\sum_{i=1}^2|\bar w_i\ra\la \bar w_i|$ is a minimizer of $a_2^*$.
We therefore conclude from \eqref{03-2} and \eqref{3.35} that \eqref{3-1} holds true. This completes the proof of   Lemma \ref{lem3.3}.\qed

\section{Mass Concentration  of Minimizers  as $a\nearrow a_2^*$}

Applying the refined estimates of the previous section, in this section we shall complete the proof of Theorem \ref{thm3} on  the concentration behavior of minimizers $\ga_a=\sum_{i=1}^2|u_i^a\ra\la u_i^a|$ for $E_a(2)$ as $a\nearrow a_2^*$, where $u_i^a\in\mathcal{H}$ satisfies \eqref{0.11} and $(u_i^a,u_j^a)=\delta_{ij}$ for $i,j=1,2$.
We start with the exponential decay of $\bar w_i^{a}(x)$ defined in \eqref{3.38} for $i=1,2$.

\begin{lem}\label{lem3.4}
Under the assumption \eqref{0-2},  suppose $\{\bar w_i^{a_n}(x)\}$ is the convergent subsequence obtained in Lemma \ref{lem3.3} (2), where $\ga_{a_n}=\sum_{i=1}^2|u_i^{a_n}\ra\la u_i^{a_n}|$ is a minimizer of $E_{a_n}(2)$ satisfying ${a_n}\nearrow a_2^*$ as $n\to\infty$. Then there exists a constant $C>0$, independent of $a_n$, such that for $i=1,2,$
\begin{equation}\label{3-4}
  |\bar w_i^{a_n}(x)|\leq C e^{-\f{\sqrt{|\lam_i|}}{2}|x|}\ \  \hbox{and}\ \ \rho_{\bar\ga_{a_n}}(x)\leq Ce^{-\sqrt{|\lam_2|}|x|}\ \  \hbox{uniformly in}\ \ \R^3
\end{equation}
as $n\to\infty$, where $\lam_i<0$ is the $i$-th eigenvalue of the operator $H_{\bar\ga}:=-\Delta-\f{5a_2^*}{3}\rho_{\bar\ga}^{\f23}$ in $\R^3$, and $\bar\ga=\sum_{i=1}^2|\bar w_i\ra\la \bar w_i|$ is as in Lemma \ref{lem3.3} (2).
\end{lem}

\ni{\bf Proof.} Since $\ga_{a_n}=\sum_{i=1}^2|u_i^{a_n}\ra\la u_i^{a_n}|$ is a minimizer of $E_{a_n}(2)$, where $u_i^{a_n}\in\mathcal{H}$ satisfies  \eqref{0.11} and $(u_i^{a_n},u_j^{a_n})=\delta_{ij}$ for $i,j=1,2$, we first claim  that
\begin{equation}\label{3-2}
  \mu_1^{a_n}<\mu_2^{a_n}<0\ \ \mbox{as}\ \ n\to\infty,
\end{equation}
where $\mu_1^{a_n}<\mu_2^{a_n}$ are the 2-first eigenvalues of the operator $-\Delta+V(x)-\f{5a_n}{3}\rho_{\ga_{a_n}}^{\f23}$ in $\R^3$, and $\rho_{\ga_{a_n}}=\sum_{i=1}^2|u_i^{a_n}|^2$.
To prove the above claim, we define
\begin{equation*}
  E_a(1)=\inf\Big\{\mathrm{Tr}\big(-\Delta+V(x)\big)\ga-a\inte \rho_\ga^{\f{5}{3}}dx:\  \ga=|u\ra\la u|,\ \|u\|_{L^2}^2=1, \ u\in\mathcal{H}\Big\},\ \ a>0.
\end{equation*}
Denote
\begin{equation*}
  a_1^*=\inf\Big\{\f{\|\ga\|^{\f23}\mathrm{Tr}(-\Delta\ga)}{\inte\rho_\ga^{\f{5}{3}}dx}:\ 0\leq\ga=\ga^*,\ \mathrm{Rank}(\ga)\leq1\Big\},
\end{equation*}
where $\rho_\ga=\beta_1|u|^2$, and $\ga=\beta_1|u\ra\la u|$ holds for $\beta_1\geq 0$ and $u\in H^1(\R^3)$.
It follows from \cite[Theorem 6]{Frank2} that $0<a_2^*<a_1^*$.
The similar argument of  \eqref{2.1} yields  that  $E_{a_n}(1)\geq0$ holds for $0<a_n<a_2^*<a_1^*$, and  hence
 \begin{equation*}
 \begin{aligned}
   0\leq E_{a_n}(1)&\leq \inte \Big(|\nabla u_1^{a_n}|^2+V(x)|u_1^{a_n}|^2\Big)dx-a_n\inte |u_1^{a_n}|^{\f{10}3}dx\\
   &=\mathrm{Tr}(-\Delta\ga_{a_n})+\inte V(x)\rho_{\ga_{a_n}}dx-a_n\inte\rho_{\ga_{a_n}}^{\f53}dx
   + a_n\inte \rho_{\ga_{a_n}}^{\f53}dx\\
   &\quad-\inte|\nabla u_2^{a_n}|^2dx-\inte V(x)|u_2^{a_n}|^2dx
   -a_n\inte \big(\rho_{\ga_{a_n}}-|u_2^{a_n}|^2\big)^{\f53}dx\\
   &=E_{a_n}(2)-\mu_2^{a_n}+ a_n\inte \rho_{\ga_{a_n}}^{\f53}dx\\
   &\quad-\f{5a_n}{3}\inte \rho_{\ga_{a_n}}^{\f23}|u_2^{a_n}|^2dx
   -a_n\inte \big(\rho_{\ga_{a_n}}-|u_2^{a_n}|^2\big)^{\f53}dx\\
 \end{aligned}
 \end{equation*}
in view of \eqref{3.15}. Applying  Lemmas \ref{lem3.1} and \ref{lem3.3}, we then deduce that
\begin{equation*}
 \begin{aligned}
   \mu_2^{a_n}\va_{a_n}^2&\leq \va_{a_n}^2E_{a_n}(2)+a_n\inte\rho_{\bar\ga_{a_n}}^{\f53}dx\\
   &\quad -\f{5a_n}{3}\inte\rho_{\bar\ga_{a_n}}^{\f23}|\bar w_2^{a_n}|^2dx-a_n\inte (\rho_{\bar\ga_{a_n}}-|\bar w_2^{a_n}|^2)^{\f53}dx\\
   &= a_2^*\inte \rho_{\bar\ga}^{\f53}dx-\f{5a_2^*}{3}\inte \rho_{\bar\ga}^{\f23}|\bar w_2|^2dx
   -a_2^*\inte \big(\rho_{\bar\ga}-|\bar w_2|^2\big)^{\f53}dx+o(1)\\
   &<0\ \ \hbox{as}\ \ n\to\infty,
 \end{aligned}
\end{equation*}
where the strict convexity of $t\mapsto t^{\f53}$ is used in the last inequality. We therefore obtain that the claim \eqref{3-2} holds true.

Following Lemma \ref{lem3.2},
we obtain  from \eqref{3.24}  that there exist constants  $C_3>0$ and $C_4>0$ such that
\begin{equation}\label{3.17}
 -C_3\leq \sum\limits_{i=1}^2\mu_i^{a_n}\va_{a_n}^2\leq-C_4\ \ \hbox{as}\ \ n\to\infty.
\end{equation}
Since $\mu_1^{a_n}<\mu_2^{a_n}<0$ as $n\to\infty$, we derive from \eqref{3.17} that $\{\mu_i^{a_n}\va_{a_n}^2\}$ is bounded uniformly as $n\to\infty$ for $i=1,2$. We thus assume that up to a subsequence if necessary,
\begin{equation}\label{3-6}
  \lim\limits_{n\to\infty}\mu_i^{a_n}\va_{a_n}^2=\lambda_i\leq 0,\ \ i=1,2.
\end{equation}
Taking the weak limit of \eqref{3.16}, we then deduce from Lemma \ref{lem3.3}  (2) that
\begin{equation*}
  -\Delta \bar w_i-\f{5a_2^*}{3}\rho_{\bar\ga}^{\f23}\bar w_i=\lambda_i \bar w_i\ \,\ \hbox{in}\ \ \R^3,\ \ i=1,2,
\end{equation*}
where $\bar w_i$ is as in \eqref{3-1} and $\rho_{\bar\ga}=\sum_{j=1}^2|\bar w_j|^2$.
Since $\bar\ga=\sum_{i=1}^2|\bar w_i\ra\la\bar w_i|$ is a minimizer of $a_2^*$, where $\bar w_i\in H^1(\R^3)$ satisfies $(\bar w_i,\bar w_j)=\delta_{ij}$ for $i,j=1,2$, one can obtain from \eqref{0.10} (or \cite[Theorem 6]{Frank2}) that $\lam_1$ and $\lam_2$ are the 2-first negative eigenvalues of the operator $H_{\bar\ga}:=-\Delta-\f{5a_2^*}{3}\rho_{\bar\ga}^{\f23}$ in $\R^3$, and hence $\lambda_1<\lambda_2<0$.

To prove \eqref{3-4}, we now establish the exponential decay of $|\bar w_i^{a_n}|$ as $n\to\infty$ for $i=1,2$.
By Kato's inequality \cite[Theorem X.27]{Reed}, we derive from \eqref{3.16} that
\begin{equation}\label{3-5}
  -\Delta|\bar w_i^{a_n}|+\Big(-\frac{5a_n}{3}\rho_{\bar\ga_{a_n}}^{\f23}-\mu_i^{a_n}\va_{a_n}^2\Big)|\bar w_i^{a_n}|\leq 0\ \ \hbox{in}\ \ \R^3,\ \ i=1,2.
\end{equation}
Because $\{\sqrt{\rho_{\bar\ga_{a_n}}}\}$ is bounded uniformly in $H^1(\R^3)$ as $n\to\infty$, by Sobolev embedding theorem, it yields  that $\{\rho_{\bar\ga_{a_n}}\}$ is bounded uniformly in $L^q(\R^3)$ as $n\to\infty$, where $1\leq q\leq 3$. We therefore obtain that  $\big\{\rho_{\bar\ga_{a_n}}^{\f23}\big\}$ is bounded uniformly in $L^r(\R^3)$ as $n\to\infty$, where $\f32\leq r\leq \f92$. Applying De Giorgi-Nash-Moser theory (cf. \cite[Theorem 4.1]{Han}), we then deduce from \eqref{3-6} and \eqref{3-5} that for any $y\in\R^3$,
\begin{equation*}
  \sup_{B_1(y)}|\bar w_i^{a_n}|\leq C\|\bar w_i^{a_n}\|_{L^2(B_2(y))}\ \ \hbox{as}\ \ n\to\infty,\ \ i=1,2,
\end{equation*}
which thus yields that for $i=1,2,$
\begin{equation*}
  |\bar w_i^{a_n}|\leq C\ \ \hbox{and}\ \ \lim\limits_{|x|\to\infty}|\bar w_i^{a_n}|=0\ \ \hbox{uniformly}\ \ \hbox{as}\ \ n\to\infty
\end{equation*}
in view of \eqref{3-1}.
This also gives that
\begin{equation}\label{3.41}
  \rho_{\bar\ga_{a_n}}\leq C\ \ \hbox{and}\ \ \lim\limits_{|x|\to\infty}\rho_{\bar\ga_{a_n}}=0\ \ \hbox{uniformly}\ \ \hbox{as}\ \ n\to\infty.
\end{equation}
Using the comparison principle, we then derive from \eqref{3-5} that
\begin{equation*}
  |\bar w_i^{a_n}|\leq C e^{-\f{\sqrt{|\lam_i|}}{2}|x|}\ \ \hbox{uniformly in}\ \ \R^3\ \ \hbox{as}\ \ n\to\infty,\ \ i=1,2,
\end{equation*}
where $\lam_1<\lam_2<0$ are the 2-first eigenvalues of the operator $H_{\bar\ga}:=-\Delta-\f{5a_2^*}{3}\rho_{\bar\ga}^{\f23}$ in $\R^3$.

To obtain the exponential decay of $\rho_{\bar\ga_{a_n}}$ as $n\to\infty$, we note from \eqref{3.16} that
for $i=1,2$,
\begin{equation*}
  -\f12\Delta|\bar w_i^{a_n}|^2+|\nabla \bar w_i^{a_n}|^2+\va_{a_n}^2 V(\va_{a_n} x+\bar x_{a_n})|\bar w_i^{a_n}|^2-\f{5a_n}{3}\rho_{\bar\ga_{a_n}}^{\f23}|\bar w_i^{a_n}|^2=\mu_i^{a_n}\va_{a_n}^2 |\bar w_i^{a_n}|^2 \,\ \ \mbox{in}\ \ \R^3,
\end{equation*}
which implies that
\begin{equation}\label{3.42}
  -\f12\Delta\rho_{\bar\ga_{a_n}}
  +\Big(-\mu_2^{a_n}\va_{a_n}^2-\f{5a_n}{3}\rho_{\bar\ga_{a_n}}^{\f23}\Big)\rho_{\bar\ga_{a_n}}\leq0\ \ \hbox{in}\ \ \R^3\ \ \hbox{as}\ \ n\to\infty.
\end{equation}
Applying the comparison principle to \eqref{3.42}, we thus obtain from \eqref{3-6} and \eqref{3.41} that
\begin{equation*}
  \rho_{\bar\ga_{a_n}}(x)\leq Ce^{-\sqrt{|\lam_2|}|x|}\ \ \hbox{uniformly in}\ \ \R^3\ \ \hbox{as}\ \ n\to\infty,
\end{equation*}
where $\lam_2<0$ is the second eigenvalue of the operator $H_{\bar\ga}$ in $\R^3$.
This completes the proof of Lemma \ref{lem3.4}.\qed


In order to prove Theorem \ref{thm3}, we next address the existence of global maximum points for $\rho_{\ga_a}(x)$, where $\ga_a=\sum_{i=1}^2|u_i^a\ra\la u_i^a|$ is a minimizer of $E_a(2)$, and $u_i^a\in\mathcal{H}$ satisfies \eqref{0.11} and  $(u_i^a,u_j^a)=\delta_{ij}$ for $i,j=1,2$. Note from \eqref{3.15} that $u_i^a$ satisfies
\begin{equation*}
  -\f12\Delta|u_i^{a}|^2+|\nabla u_i^a|^2+V(x)|u_i^a|^2-\f{5a}{3}\rho_{\ga_a}^{\f23}|u_i^a|^2=\mu_i^a|u_i^a|^2\ \ \hbox{in}\ \ \R^3,\ \ i=1,2.
\end{equation*}
We therefore obtain that
\begin{equation}\label{4.7}
  -\f12\Delta\rho_{\ga_a}+\Big(-\f{5a}{3}\rho_{\ga_a}^{\f23}
  -\mu_2^a\Big)\rho_{\ga_a}\leq0\ \ \hbox{in}\ \ \R^3,
\end{equation}
due to the fact that $\mu_1^a<\mu_2^a$.
Since $u_i^a\in H^1(\R^3)$ for $i=1,2$, it yields from \eqref{3.53} that $\sqrt{\rho_{\ga_a}}\in H^1(\R^3)$. By Sobolev's embedding theorem, it then gives  that $\rho_{\ga_{a}}\in L^q(\R^3)$ for  $1\leq q\leq 3$, and hence $\rho_{\ga_{a}}^{\f23}\in L^r(\R^3)$ for $\f32\leq r\leq \f92$.
Following De Giorgi-Nash-Moser theory (cf. \cite[Theorem 4.1]{Han}), we then obtain from \eqref{4.7} that for any $y\in \R^3$,
\begin{equation*}
  \sup_{B_1(y)}\rho_{\ga_a}(x)\leq C\|\rho_{\ga_a}\|_{L^1(B_2(y))},
\end{equation*}
which yields that $\lim_{|x|\to\infty}\rho_{\ga_a}(x)=0$. Because $\inte \rho_{\ga_a}(x)dx=2$, this further gives that
global maximum points of $\rho_{\ga_a}(x)$ exist in a bounded ball $B_R(0)$, where $R>0$ is large enough.

Applying the above existence of global maximum points for $\rho_{\ga_a}(x)$, we next analyze the following convergence.

\begin{lem}\label{lem4.1}
Under the assumption \eqref{0-2}, assume the constant $p>0$ and the set $\Lambda$ are defined by \eqref{0-4}. Suppose $\{\bar w_i^{a_n}(x)\}$ is the convergent subsequence obtained in Lemma \ref{lem3.3} (2), where $\ga_{a_n}=\sum_{i=1}^2|u_i^{a_n}\ra\la u_i^{a_n}|$ is a minimizer of $E_{a_n}(2)$ satisfying ${a_n}\nearrow a_2^*$ as $n\to\infty$. Then up to a subsequence if necessary,
\begin{equation}\label{4.1}
w_i^{a_n}(x):=\va_{a_n}^{\f32}u_i^{a_n}(\va_{a_n} x+ x_{a_n})\stackrel{n}{\longrightarrow} w_i(x), \ \ \va_{a_n}:=(a_2^*-a_n)^{\f{1}{p+2}}>0
\end{equation}
strongly in $H^1(\R^3)\cap L^\infty(\R^3)$ for $i=1,2$, where $\ga:=\sum_{i=1}^2|w_i\ra\la w_i|$ is a minimizer of $a_2^*$ defined by \eqref{0.7}, and there exists a point $x_k\in\Lambda$ such that the global maximum point $x_{a_n}\in\R^3$ of $\rho_{\ga_{a_n}}(x)=\sum_{i=1}^2|u_i^{a_n}|^2$ satisfies
\begin{equation}\label{4.1M}
  x_{a_n}\stackrel{}{\longrightarrow} x_k \ \ as \ \ n\to\infty.
\end{equation}
\end{lem}

\ni{\bf Proof.}
Define for $i=1,2,$
\begin{equation}\label{3.44}
   w_i^{a_n}(x):=\va_{a_n}^{\f32}u_i^{a_n}(\va_{a_n} x+x_{a_n})=\bar w_i^{a_n}\Big(x+\f{x_{a_n}-\bar x_{a_n}}{\va_{a_n}}\Big),\ \ \va_{a_n}:=(a_2^*-a_n)^{\f{1}{p+2}}>0,
\end{equation}
and
\begin{equation}\label{3-7}
  \hat\ga_{a_n}:=\sum\limits_{i=1}^2| w_i^{a_n}\ra\la w_i^{a_n}|,
\end{equation}
where  $\bar w_i^{a_n}(x)$ and $\bar x_{a_n}\in\R^3$ are as in  \eqref{3-1}, and $x_{a_n}\in\R^3$ is a global maximum point of $\rho_{\ga_{a_n}}(x)=\sum_{i=1}^2|u_i^{a_n}|^2$.
It then follows from \eqref{3.15} and \eqref{3.44} that $ w_i^{a_n}(x)$ satisfies the following  system
\begin{equation}\label{3.49}
  -\Delta w_i^{a_n}+\va_{a_n}^2 V(\va_{a_n} x+x_{a_n}) w_i^{a_n}-\f{5a_n}{3}\rho_{\hat\ga_{a_n}}^{\f23} w_i^{a_n}=\mu_i^{a_n}\va_{a_n}^2w_i^{a_n}\ \ \hbox{in}\ \ \R^3,\ \ i=1,2,
\end{equation}
where $\rho_{\hat\ga_{a_n}}=\sum_{j=1}^2|w_j^{a_n}|^2$, and $\mu_1^{a_n}<\mu_2^{a_n}$ are the 2-first eigenvalues of the operator $-\Delta+V(x)-\f{5a_n}{3}\rho_{\ga_{a_n}}^{\f23}$ in $\R^3$.


We first claim that there exists a constant $C>0$, independent of $a_n>0$, such that
\begin{equation}\label{3.47}
  \f{|x_{a_n}-\bar x_{a_n}|}{\va_{a_n}}\leq C\ \ \hbox{uniformly as}\ \ n\to\infty.
\end{equation}
In fact, if \eqref{3.47} is false, then there exists a subsequence, still denoted by $\{a_n\}$, of $\{a_n\}$ such that $\f{|x_{a_n}-\bar x_{a_n}|}{\va_{a_n}}\to\infty$ as $n\to\infty$. It thus follows from
\eqref{3-4} that
\begin{equation}\label{3.54}
  \rho_{\ga_{a_n}}(x_{a_n})=\va_{a_n}^{-3}\rho_{\bar\ga_{a_n}}\Big(\f{x_{a_n}-\bar x_{a_n}}{\va_{a_n}}\Big)\leq C\va_{a_n}^{-3}e^{-\f{\sqrt{|\lam_2|}|x_{a_n}-\bar x_{a_n}|}{\va_{a_n}}}=o(\va_{a_n}^{-3})\ \ \hbox{as}\ \ n\to\infty,
\end{equation}
where $\bar\ga_{a_n}$ is as in \eqref{3.38}.
On the other hand, it follows from \eqref{4.7} that $\rho_{\ga_{a_n}}(x)=\sum_{i=1}^2|u_i^{a_n}|^2$ satisfies
\begin{equation*}
  -\f12\Delta\rho_{\ga_{a_n}}(x)-\f{5a_n}{3}\rho_{\ga_{a_n}}^{\f53}(x)
  \leq \mu_2^{a_n}\rho_{\ga_{a_n}}(x)\ \,\ \hbox{in}\ \ \R^3.
\end{equation*}
Since $x_{a_n}\in\R^3$ is a maximum point of $\rho_{\ga_{a_n}}(x)$, we have $-\f12\Delta\rho_{\ga_{a_n}}(x_{a_n})\geq 0$, and hence
\begin{equation*}
  \rho_{\ga_{a_n}}(x_{a_n})\geq \Big(\f{-3\mu_2^{a_n}}{5a_n}\Big)^{\f32}\geq C\va_{a_n}^{-3}\ \ \hbox{as}\ \ n\to\infty,
\end{equation*}
due to the fact that $\lim_{n\to\infty}\mu_2^{a_n}\va_{a_n}^2=\lam_2<0$. This however contradicts with \eqref{3.54}. We therefore derive that the claim \eqref{3.47} holds true.

Applying \eqref{3.47}, then there exists a constant $R_1>0$, independent of $a_n$, such that $\f{|x_{a_n}-\bar x_{a_n}|}{\va_{a_n}}<\f{R_1}{2}$ as $n\to\infty$. Moreover, it then yields from \eqref{3-1} that there exists a point $x_k\in\Lambda$ such that the maximum point $x_{a_n}$ of $\rho_{\ga_{a_n}}(x)$ satisfies
 \begin{equation}\label{4.8}
   \lim\limits_{n\to\infty}x_{a_n}=\lim\limits_{n\to\infty}\bar x_{a_n}=x_k\in\Lambda,
 \end{equation}
which thus proves (\ref{4.1M}).
Following \eqref{3.44}, we have
 \begin{equation*}
   \rho_{\hat\ga_{a_n}}(x)=\rho_{\bar\ga_{a_n}}\Big(x+\f{x_{a_n}-\bar x_{a_n}}{\va_{a_n}}\Big),
 \end{equation*}
 where $\rho_{\hat\ga_{a_n}}(x)=\sum_{i=1}^2|w_i^{a_n}|^2$ and $\rho_{\bar\ga_{a_n}}(x)=\sum_{i=1}^2|\bar w_i^{a_n}|^2$.
 It then follows from \eqref{3.8} that
 \begin{equation*}
 \begin{aligned}
   \lim\limits_{n\to\infty}\int_{B_{R_0+R_1}(0)} \rho_{\hat\ga_{a_n}}(x)dx
   &=\lim\limits_{n\to\infty}\int_{B_{R_0+R_1}(\f{x_{a_n}-\bar x_{a_n}}{\va_{a_n}})}\rho_{\bar\ga_{a_n}}(x)dx\\
   &\geq\lim\limits_{n\to\infty}\int_{B_{R_0}(0)}\rho_{\bar\ga_{a_n}}(x)dx\geq\eta>0.
 \end{aligned}
 \end{equation*}
 The similar argument of proving  \eqref{3-1} thus yields that there exist a subsequence, still denoted by $\{a_n\}$, of $\{a_n\}$ and $w_i\in H^1(\R^3)$ such that for $ i=1,2,$
\begin{equation}\label{3.45}
  w_i^{a_n}(x):=\va_{a_n}^{\f32}u_i^{a_n}(\va_{a_n} x+x_{a_n})\to   w_i(x)\ \ \hbox{strongly in}\ \ H^1(\R^3)\ \ \hbox{as}\ \ n\to\infty,\ \
\end{equation}
where
$\ga:=\sum_{i=1}^2|  w_i\ra\la  w_i|$ is a minimizer of $a_2^*$ defined by \eqref{0.7}.

We next prove (\ref{4.1}) on the $L^\infty$-uniform convergence of $w_i^{a_n}(x)$ as $n\to\infty$. Similar to Lemmas \ref{lemA.1} and \ref{lem3.4}, one can derive that
\begin{equation}\label{4.10}
  |  w_i(x)|,\ |  w_i^{a_n}(x)|\leq C e^{-\f{\sqrt{|\mu_i|}}{2}|x|}\ \ \hbox{uniformly in}\ \ \R^3\ \ \hbox{as}\ \ n\to\infty,\ \ i=1,2,
\end{equation}
where $\mu_1<\mu_2<0$ are the 2-first eigenvalues of the operator  $H_{\ga}:=-\Delta-\f{5a_2^*}{3}\rho_{\ga}^{\f23}$ in $\R^3$.
On the other hand, define
\begin{equation*}
  G_i^{a_n}(x):=-\va_{a_n}^2 V(\va_{a_n} x+ x_{a_n}) w_i^{a_n}(x)+\f{5a_n}{3}\rho_{\hat\ga_{a_n}}^{\f23} w_i^{a_n}(x)+\mu_i^{a_n}\va_{a_n}^2 w_i^{a_n}(x),\ \ i=1,2,
\end{equation*}
so that the system \eqref{3.49} can be rewritten as
\begin{equation}\label{3.52}
  -\Delta  w_i^{a_n}(x)= G_i^{a_n}(x)\ \ \hbox{in}\ \ \R^3,\ \ i=1,2.
\end{equation}
Since it follows from \eqref{4.10} that $\{ w_i^{a_n}\}$ and $\{\rho_{\hat\ga_{a_n}}\}$ are bounded uniformly in $L^\infty(\R^3)$ as $n\to\infty$,  we deduce from \eqref{0-2},  \eqref{3-6}  and \eqref{4.8} that $\{G_i^{a_n}\}$ is bounded uniformly in $L^p_{loc}(\R^3)$ for $p>2$ as $n\to\infty$. Applying the $L^p$ theory to \eqref{3.52}, it further yields that $\{w_i^{a_n}\}$ is bounded uniformly in $W^{2,p}_{loc}(\R^3)$ as $n\to\infty$.  We therefore obtain from \cite[Theorem 7.26]{GT} that there exist a subsequence, still denoted by $\{ w_i^{a_n}\}$,  of $\{ w_i^{a_n}\}$ and $\hat w_i(x)$ such that
\begin{equation*}
   w_i^{a_n}(x)\to \hat w_i(x)\ \ \hbox{uniformly in}\ \ L^\infty_{loc}(\R^3)\ \ \hbox{as}\ \ n\to\infty,\ \ i=1,2.
\end{equation*}
Note from \eqref{3.45} that $\hat w_i(x)=  w_i(x)$, and hence
\begin{equation}\label{4.9}
   w_i^{a_n}(x):=\va_{a_n}^{\f32}u_i^{a_n}(\va_{a_n} x+x_{a_n})\to   w_i(x)\ \ \hbox{uniformly in}\ \ L^\infty_{loc}(\R^3)\ \ \hbox{as}\ \ n\to\infty,\ \ i=1,2.
\end{equation}
We thus conclude from \eqref{4.8}--\eqref{4.10} and \eqref{4.9} that  the $L^\infty$-uniform convergence (\ref{4.1})  holds true, which therefore completes the proof of Lemma  \ref{lem4.1}.\qed

Applying Lemma  \ref{lem4.1}, we are now ready to establish Theorem \ref{thm3}.

\vskip 0.05truein
\noindent{\bf Proof of Theorem \ref{thm3}.}
In view of Lemma \ref{lem4.1}, to complete the proof of Theorem \ref{thm3}, it suffices to prove that the point $x_k$ of \eqref{4.1M} satisfies
\begin{equation}\label{4.6}
 x_k\in\mathcal{Z}\ \ \hbox{and}\ \  
 \lim\limits_{n\to\infty}\f{x_{a_n}-x_k}{\va_{a_n}}=\bar x, \ \ \va_{a_n}=(a_2^*-a_n)^{\f{1}{p+2}}>0,
\end{equation}
where the set $\mathcal{Z}$ is defined by \eqref{0-3}, $\bar x$ is some point in $\R^3$, and $x_{a_n}\in\R^3$ is a maximum point of $\rho_{\ga_{a_n}}(x)=\sum_{i=1}^2|u_i^{a_n}|^2$.
By direct calculations, we deduce from \eqref{0.7}  and \eqref{4.1} that
\begin{equation}\label{4.3}
\begin{aligned}
  E_{a_n}(2)&=\mathrm{Tr}\big(-\Delta+V(x)\big)\ga_{a_n}-a_n\inte\rho_{\ga_{a_n}}^{\f53}dx\\
  &=\va_{a_n}^{-2}\Big(\mathrm{Tr}(-\Delta\hat\ga_{a_n})
  -a_2^*\inte\rho_{\hat\ga_{a_n}}^{\f53}dx\Big)\\
&\quad+\inte V(\va_{a_n} x+x_{a_n})\rho_{\hat\ga_{a_n}}dx
+\va_{a_n}^p\inte\rho_{\hat\ga_{a_n}}^{\f53}dx\\
&\geq \inte V(\va_{a_n} x+x_{a_n})\rho_{\hat\ga_{a_n}}dx
+\va_{a_n}^p\inte\rho_{\hat\ga_{a_n}}^{\f53}dx,\\
\end{aligned}
\end{equation}
where $\rho_{\hat\ga_{a_n}}=\sum_{i=1}^2|w_i^{a_n}|^2$, and $\hat\ga_{a_n}=\sum_{i=1}^2|w_i^{a_n}\ra\la w_i^{a_n}|$ is defined by \eqref{3-7}.

We now claim that
\begin{equation}\label{4.2}
  \Big\{\frac{|x_{a_n}-x_k|}{\va_{a_n}}\Big\}\ \ \hbox{is bounded uniformly as}\ \  n\to\infty.
\end{equation}
On the contrary, assume that \eqref{4.2} is false. We then obtain that there exists a subsequence, still denoted by $\{a_n\}$, of $\{a_n\}$ such that
\begin{equation*}
  \lim\limits_{n\to\infty}\f{|x_{a_n}-x_k|}{\va_{a_n}}=\infty.
\end{equation*}
It thus follows from Fatou's lemma that for any sufficiently large $M'>0$,
\begin{equation}\label{4.14}
\begin{aligned}
  &\liminf_{n\to\infty}\va_{a_n}^{-p_k}\inte V(\va_{a_n}x+x_{a_n})\rho_{\hat\ga_{a_n}}dx\\
  =&\liminf_{n\to\infty}\inte \f{V(\va_{a_n}x+x_{a_n})}{|\va_{a_n}x+x_{a_n}-x_k|^{p_k}}\Big|x
  +\f{x_{a_n}-x_k}{\va_{a_n}}\Big|^{p_k}\rho_{\hat\ga_{a_n}}dx\\
  \geq&\inte \liminf_{n\to\infty} \f{V(\va_{a_n}x+x_{a_n})}{|\va_{a_n}x+x_{a_n}-x_k|^{p_k}}\Big|x
  +\f{x_{a_n}-x_k}{\va_{a_n}}\Big|^{p_k}\rho_{\hat\ga_{a_n}}dx
  \geq M',
\end{aligned}
\end{equation}
where $p_k>0$ is as in \eqref{0-2}.
We further derive from \eqref{3.7}, \eqref{4.3} and \eqref{4.14} that
\begin{equation}\label{4.15}
  E_{a_n}(2)\geq \f{M'}{2}\va_{a_n}^{p_k}=\f{M'}{2}(a_2^*-a_n)^{\f{p_k}{p+2}}\ \ \hbox{as}\ \ n\to\infty
\end{equation}
holds for above any constant $M'>0$, which however contradicts with Lemma \ref{lem3.1}. We therefore conclude that the claim \eqref{4.2} holds true. The same argument of \eqref{4.14} and \eqref{4.15} also yields that $p_k=p$.

It follows from the claim \eqref{4.2} that there exist a subsequence, still denoted by $\{a_n\}$, of $\{a_n\}$ and a point $\bar x\in\R^3$ such that
\begin{equation}\label{4.13}
  \lim\limits_{n\to\infty}\f{x_{a_n}-x_k}{\va_{a_n}}=\bar x.
\end{equation}
We then obtain from Lemma  \ref{lem4.1} and \eqref{0-5} that
\begin{equation}\label{4.5}
\begin{aligned}
  &\liminf_{n\to\infty}\va_{a_n}^{-p}\inte V(\va_{a_n}x+x_{a_n})\rho_{\hat\ga_{a_n}}dx\\
  =&\liminf_{n\to\infty}\inte \f{V(\va_{a_n}x+x_{a_n})}{|\va_{a_n}x+x_{a_n}-x_k|^{p}}\Big|x
  +\f{x_{a_n}-x_k}{\va_{a_n}}\Big|^{p}\rho_{\hat\ga_{a_n}}dx\\
  \geq&\al_k\inte |x+\bar x|^{p}\rho_\ga dx\geq\al\inte |x+\bar x|^{p}\rho_\ga dx,
\end{aligned}
\end{equation}
where $\ga=\sum_{i=1}^2|w_i\ra\la w_i|$ is as in Lemma \ref{lem4.1}, and all above identities hold, if and only if  $\al_k=\al$ is as in \eqref{0-5}.
We thus deduce from \eqref{4.3} and \eqref{4.5} that
\begin{equation}\label{4.11}
\begin{aligned}
  \liminf_{n\to\infty}\f{E_{a_n}(2)}{\va_{a_n}^p}
  &\geq \inte\rho_{\ga}^{\f53}dx+\al\inte |x+\bar x|^{p}\rho_\ga dx.
\end{aligned}
\end{equation}

On the other hand, defining
\begin{equation*}
  u_i(x)=\va_{a_n}^{-\f32}w_{i}\Big(\f{x-x_m}{\va_{a_n}}-\bar x\Big),\ \ i=1,2,
\end{equation*}
where $x_m\in\mathcal{Z}$ is as in \eqref{0-3},
choose $\ga_1=\sum_{i=1}^2|u_i\ra\la u_i|$ as a trail operator of $E_{a_n}(2)$, and assume $\ga=\sum_{i=1}^2|w_i\ra\la w_i|$ defined in Lemma \ref{lem4.1} is a minimizer of $a_2^*$ and $\|\ga\|=1$. We then deduce from \eqref{3.7} that
\begin{equation*}
\begin{aligned}
  E_{a_n}(2)&\leq \mathrm{Tr}\big(-\Delta+V(x)\big)\ga_1-a_n\inte\rho_{\ga_1}^{\f53}dx\\
  &=\va_{a_n}^{-2}\Big(\mathrm{Tr}(-\Delta\ga)-a_n\inte\rho_{\ga}^{\f53}dx\Big)
+\inte V\big(\va_{a_n}(x+\bar x)+x_m\big)\rho_\ga dx\\
&=\va_{a_n}^p\Big\{\inte\rho_{\ga}^{\f53}dx+\inte\f{V\big(\va_{a_n}(x+\bar x)+x_m\big)}{|\va_{a_n}(x+\bar x)+x_m-x_m|^p}|x+\bar x|^p\rho_\ga dx\Big\},
\end{aligned}
\end{equation*}
which yields that
\begin{equation}\label{4.12}
  \limsup_{n\to\infty}\f{E_{a_n}(2)}{\va_{a_n}^p}\leq \inte\rho_{\ga}^{\f53}dx+\al \inte|x+\bar x|^p\rho_{\ga}dx.
\end{equation}
We thus conclude from \eqref{4.11} and \eqref{4.12} that
\begin{equation*}
  \lim\limits_{n\to\infty}\f{E_{a_n}(2)}{\va_{a_n}^p}= \inte\rho_{\ga}^{\f53}dx+\al \inte|x+\bar x|^p\rho_{\ga}dx.
\end{equation*}
Together with  \eqref{4.5}, this further implies that $\al_k=\al$, and hence \eqref{4.6} holds true. This completes the proof of Theorem \ref{thm3}.\qed

%
%

\appendix
\section{Appendix}
For the reader's convenience, the purpose of  this appendix is to establish  the following exponential decay of minimizers for $a_2^*$.
\begin{lem}\label{lemA.1}
Assume
\begin{equation}\label{2.15}
  \ga^{(2)}=\|\ga^{(2)}\|\sum\limits_{i=1}^{2}|Q_i\ra\la Q_i|,\ \ Q_i\in H^1(\R^3),\ \  (Q_i,Q_j)=\delta_{ij},\ \ i,j=1,2,
\end{equation}
is a minimizer of $a_2^*$ defined by \eqref{0.7}. Then we have
\begin{equation*}
 Q_i\in C^2(\R^3),\ \  |Q_i|\leq Ce^{-\f{\sqrt{|\hat\mu_i|}}2|x|}\ \ \hbox{and}\ \  |\nabla Q_i|\leq Ce^{-\f{\sqrt{|\hat\mu_i|}}4|x|}\ \ \hbox{in}\ \ \R^3,\ \ i=1,2,
\end{equation*}
where $\hat\mu_1<\hat\mu_2<0$ are the 2-first negative eigenvalues of the operator
\begin{equation}\label{A.1}
\hat H_{\ga}:=-\Delta-\f53 a_2^*\rho_{\ga}^{\f23}\ \ \hbox{in}\ \ \R^3,\ \   \rho_{\ga}=\sum\limits_{j=1}^{2}|Q_j|^2  \ \ \hbox{and}\ \ \ga:=\sum_{i=1}^2|Q_i\ra\la Q_i|.
\end{equation}
\end{lem}

\ni{\bf Proof.}
Since $\ga^{(2)}$ is a minimizer of $a_2^*$, $Q_i(x)$ satisfies the following system
\begin{equation}\label{2.4}
  -\Delta Q_i(x)-\f53 a_2^*\rho_{\ga}^{\f23}Q_i(x)=\hat\mu_i Q_i(x)\ \ \hbox{in}\ \ \R^3,\ \ i=1,2,
\end{equation}
where $\hat\mu_1<\hat\mu_2<0$ are the 2-first negative eigenvalues of the operator $\hat H_{\ga}$ defined in \eqref{A.1}.
We first claim that
\begin{equation}\label{2.5}
  Q_i(x)\in C^2(\R^3)\ \ \hbox{and}\ \ \lim\limits_{|x|\to\infty}|Q_i(x)|=0,\ \ i=1,2.
\end{equation}
In fact, by Kato's inequality (cf. \cite[Theorem X.27]{Reed}), we derive from \eqref{2.4} that
\begin{equation}\label{2.6}
 -\Delta|Q_i|+\Big(-\f53 a_2^*\rho_{\ga}^{\f23}-\hat\mu_i\Big)|Q_i|\leq 0\,\ \ \hbox{in}\ \ \R^3,\ \ i=1,2.
\end{equation}
Since $Q_i(x)\in H^1(\R^3)$ for $i=1,2$, we have $\rho_{\ga}(x)\in L^q(\R^3)$ for $1\leq q\leq 3$, and hence $\rho_{\ga}^{\f23}(x)\in L^r(\R^3)$ for $\f32\leq r\leq \f92$. Applying De Giorgi-Nash-Moser
theory (cf. \cite[Theorem 4.1]{Han}), it then yields from \eqref{2.6} that for any $y\in\R^3$,
\begin{equation*}
  \sup_{B_1(y)}|Q_i|\leq C\|Q_i\|_{L^2(B_2(y))},\ \ i=1,2,
\end{equation*}
which implies that $Q_i(x)\in L^\infty(\R^3)$ and $\lim\limits_{|x|\to\infty}|Q_i|=0$ for $i=1,2$. This also gives that $\rho_{\ga}(x)\in L^\infty(\R^3)$ and $\lim\limits_{|x|\to\infty}\rho_{\ga}(x)=0$.

We next  prove the continuity of $Q_i(x)$ for $i=1,2$. Denoting
 $$G_i(x):=\Big(\f53 a_2^*\rho_{\ga}^{\f23}+\hat\mu_i\Big)Q_i(x),$$
we obtain from \eqref{2.4} that
\begin{equation}\label{2.8}
  -\Delta Q_i(x)=G_i(x)\ \ \hbox{in}\ \ \R^3,\ \ i=1,2.
\end{equation}
Since $Q_i(x)\in L^\infty(\R^3)$, we derive that $G_i(x)\in L^q_{loc}(\R^3)$ holds for $q> 2$. Applying the $ L^p$ theory (cf. \cite[Theorem 9.11]{GT}), we then deduce from \eqref{2.8} that $Q_i(x)\in W^{2,q}_{loc}(\R^3)$ for $i=1,2$. The standard Sobolev
embedding theorem thus gives that $Q_i(x)\in C^\theta_{loc}(\R^3)$ holds  for some $\theta\in (0,1)$. By the Schauder estimate (cf. \cite[Theorem
6.2]{GT}), we further obtain that $Q_i\in C^{2,\theta}_{loc}(\R^3)$, and hence $Q_i(x)\in C^2(\R^3)$ for $i=1,2$.  This gives the proof of \eqref{2.5}.

We finally prove the exponential decay of $|Q_i|$ for $i=1,2$.
Since $\lim_{|x|\to\infty}\rho_{\ga}(x)=0$,
applying the comparison principle, it gives from \eqref{2.6} that there exists a constant $C>0$ such that
\begin{equation}\label{2.12}
  |Q_i|\leq Ce^{-\f{\sqrt{|\hat\mu_i|}}2|x|}\ \,\ \hbox{in}\ \ \R^3,\ \ i=1,2.
\end{equation}
By gradient estimates of (3.15) in \cite{GT}, we further derive from  \eqref{2.4} and \eqref{2.12} that
$$|\nabla Q_i|\leq Ce^{-\f{\sqrt{|\hat\mu_i|}}4|x|}\ \ \hbox{in}\ \ \R^3,\ \ i=1,2,$$
which therefore completes the proof of Lemma \ref{lemA.1}.\qed

\end{document}